\documentclass{jfm}

\usepackage{color}

\begin{document}

\newtheorem{lemma}{Lemma}
\newtheorem{corollary}{Corollary}

\providecommand\Gr{\mbox{\it{Gr}} }
\providecommand\Sc{\mbox{\it{Sc}} }
\providecommand\Pe{\mbox{\it{Pe}} }
\providecommand\Rey{\mbox{\it{Re}} }
\providecommand\Ra{\mbox{\it{Ra}} }
\providecommand\Ja{\mbox{\it{Ja}} }
\providecommand\tinabla{\tilde\nabla}
\providecommand\btinabla{\boldsymbol{\tilde\nabla}}
\providecommand\bcdot{\boldsymbol{\cdot}}
\providecommand\bU{\boldsymbol{U}}
\providecommand\bOmega{\boldsymbol{\Omega}}
\providecommand\bu{\boldsymbol{u}}
\providecommand\bv{\boldsymbol{v}}
\providecommand\bg{\boldsymbol{g}}
\providecommand\binfty{\boldsymbol{\infty}}
\providecommand\ghat{\boldsymbol{\hat g}}
\providecommand\eeta{\boldsymbol{\hat e}_{\boldsymbol \eta}}
\providecommand\exi{\boldsymbol{\hat e}_{\boldsymbol \xi}}
\providecommand\ephi{\boldsymbol{\hat e}_{\boldsymbol \phi}}
\providecommand\nhat{\boldsymbol{\hat n}}
\providecommand\bx{\boldsymbol{x}}
\providecommand\ex{\boldsymbol{\hat e}_{\boldsymbol x}}
\providecommand\ey{\boldsymbol{\hat e}_{\boldsymbol y}}
\providecommand\ez{\boldsymbol{\hat e}_{\boldsymbol z}}
\providecommand\xad{\tilde{x}}
\providecommand\yad{\tilde{y}}
\providecommand\zad{\tilde{z}}
\providecommand\denh{\cosh(\theta\eta) - \cos(\theta\xi)}
\providecommand\sech{\mathrm{sech}}
\providecommand\hti{\tilde h}
\providecommand\htiphi{\tilde h_{\phi}}
\providecommand\ttau{\tilde\tau}
\providecommand\dd{\mathrm{d}}
\providecommand\dD{\mathrm{D}}
\providecommand\den{\eta^2 + \xi^2}
\providecommand\Vc{\overline V^\infty_g}
\providecommand\cdg{\mbox{CO$_2$} }
\providecommand\cdaq{\mbox{CO$_\mathrm{2\:(aq)}$} }
\providecommand\hco{\mbox{HCO$_3^-$} }
\providecommand\Rcorr{R_\mathit{corr}}
\providecommand\gra{\lambda|\Delta C|g}

\shorttitle{The history effect on bubble growth and dissolution. Part 2} 

\shortauthor{P. Pe\~nas-L\'opez, \'A. Moreno Soto et al} 

\title{The history effect on bubble growth and dissolution. Part 2. Experiments and simulations of a spherical bubble attached to a horizontal flat plate}

\author
 {
 Pablo Pe\~nas-L\'opez\aff{1}
 \corresp{\email{papenasl@ing.uc3m.es}},
 \'Alvaro Moreno Soto\aff{2}
 \corresp{\email{a.morenosoto@utwente.nl}},
 Miguel A. Parrales\aff{3},
 Devaraj van der Meer\aff{2},
 Detlef Lohse\aff{2}
 \and
 Javier Rodr\'{\i}guez-Rodr\'{\i}guez\aff{1}
 }

\affiliation
{
\aff{1}
Fluid Mechanics Group, Universidad Carlos III de Madrid, Avda. de la Universidad 30, 28911 Legan\'es (Madrid), Spain

\aff{2}
Physics of Fluids Group, Faculty of Science and Technology, University of Twente, P.O. Box 217, 7500 AE Enschede, The Netherlands

\aff{3}
Departamento de Ingenier\'{\i}a Energ\'etica y Fluidomec\'anica, Escuela T\'ecnica Superior de Ingenieros Industriales, Universidad Polit\'ecnica de Madrid, C. Jos\'e Guti\'errez Abascal, 2. 28006, Madrid, Spain
}

\maketitle

\begin{abstract}
The accurate description of the growth or dissolution dynamics of a soluble gas bubble in a super- or undersaturated solution requires taking into account a number of physical effects that contribute to the instantaneous mass transfer rate. One of these effects is the so-called history effect. It refers to the contribution of the local concentration boundary layer around the bubble that has developed from past mass transfer events between the bubble and liquid surroundings.
In Part 1 of this work \citep{penas2016}, a theoretical treatment of this effect was given for a spherical, isolated bubble. Here, Part 2 provides an experimental and numerical study of the history effect regarding a spherical bubble attached to a horizontal flat plate and in the presence of gravity. The simulation technique developed in this paper is based on a streamfunction--vorticity formulation that may be applied to other flows where bubbles or drops exchange mass in the presence of a gravity field.
Using this numerical tool, simulations are performed for the same conditions used in the experiments, in which the bubble is exposed to subsequent growth and dissolution stages, using step-wise variations in the ambient pressure. Besides proving the relevance of the history effect, the simulations highlight the importance that boundary-induced advection has to accurately describe bubble growth and shrinkage, i.e. the bubble radius evolution. In addition, natural convection has a significant influence that shows up in the velocity field even at short times, though, given the supersaturation conditions studied here, the bubble evolution is expected to be mainly diffusive.
\end{abstract}

\section{Introduction} 
\label{sec:intro}
Mass transfer processes involving bubbles have gained a renewed interest over the last few years due to their relevance in modern microfluidic applications, connected to topics such as carbon sequestration \citep{sun2011, volk2015}. Due to the small size of these bubbles, they are  spherical once they become smaller than the channel's size and are detached from the channel's wall. Thus, in general terms, the theory of \cite{epstein1950} describing the diffusion-driven growth or dissolution of an isolated, spherical particle should be applicable. However, as discussed in Part 1 of this work \citep{penas2016}, a number of effects not included in the Epstein--Plesset theory, e.g. flow around the bubble, must be taken into account to properly describe various experimental observations. Bubbles may also interact with nearby surfaces or they may contain more than one chemical species \citep{shim2014, penas2015}. Another effect that contributes to the diffusion-driven dynamics of a bubble is the so-called history effect, discussed in Part 1 and more recently in \citet{chu2016.1}. It has been shown that any recent history of growth and/or dissolution (triggered by past changes in ambient pressure) experienced by a particular bubble may leave, at least for some time, a non-negligible print on the current state of the concentration profile surrounding such bubble. Consequently, the mass transfer rate is affected as well. In Part 1, we proposed a modification to the theory of Epstein \& Plesset to take into account the history effect through a memory integral term for the case of spherical, isolated bubbles. Moreover, we applied this modified equation to calculate the bubble radius evolution when the bubble is subjected to some simple, yet relevant, pressure-time histories. It is worth mentioning that history effects are common to problems in which diffusion plays a central role, such as the viscous drag around a body or, closer to the present mass-transfer problem, the heat transfer around a spheroid \citep{michaelides2003}.

\begin{figure}
  \centerline{\includegraphics[width=0.9\textwidth]{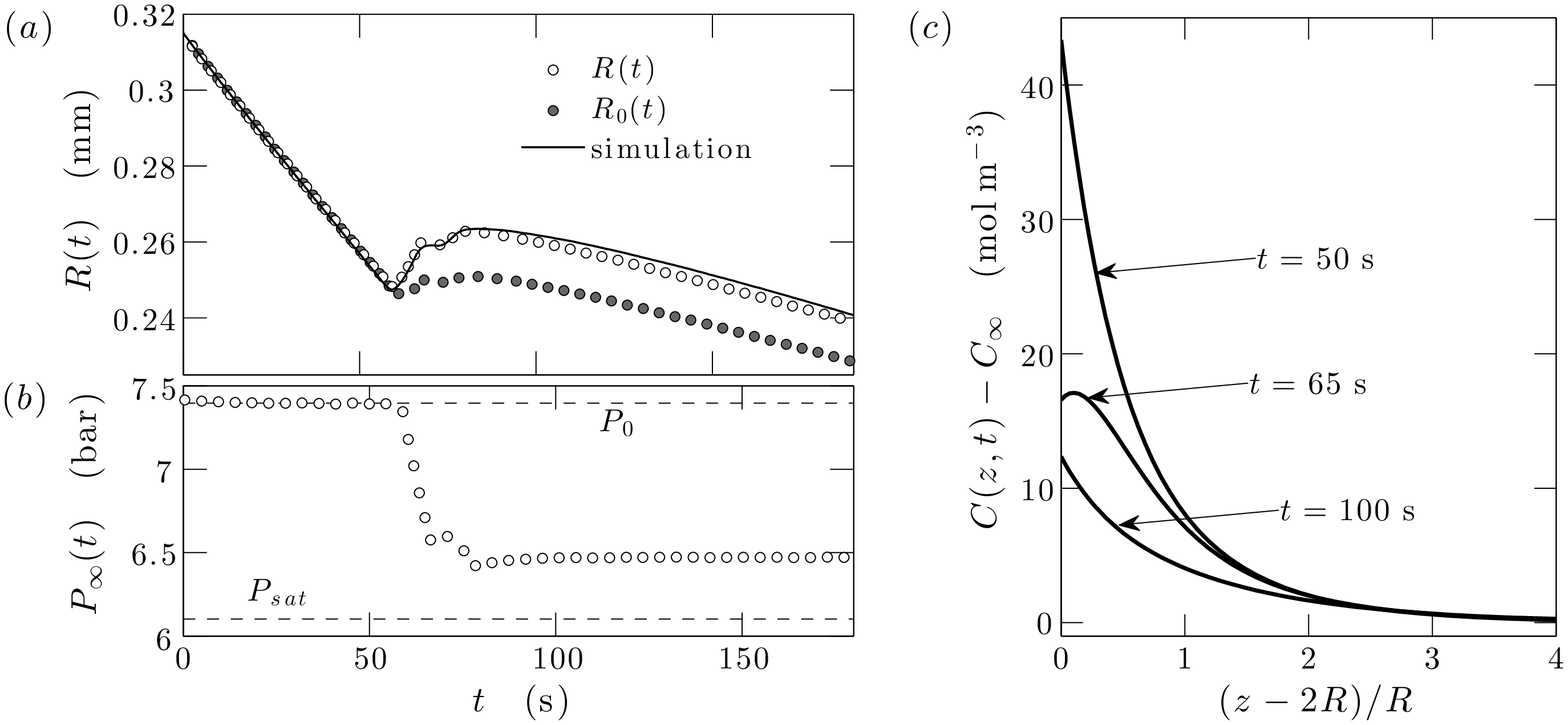}}
  \caption{\label{fig:intro_s0}Dissolution of a CO$_2$ spherical-cap bubble tangent to a flat chip immersed in a CO$_2$-water solution under pressurised conditions (see later figure \ref{fig:formulation}). The bubble is subjected to (b) a pressure jump $P_\infty(t)$, from $P_\infty(0)=P_0 = 7.4$ bar to 6.5 bar. Both pressures are above the saturation pressure, $P_\mathit{sat} = 6.1$ bar (according to simulation). Panel (a) shows the evolution in time of the measured bubble radius $R(t)$ (white markers) and ambient radius $R_0(t)$ (dark markers). The former is compared to simulation, which in addition was employed to depict (c) the concentration profile along the $z$-axis above the bubble at three different instants in time. The employed experimental and numerical techniques are detailed in the main text.}
\end{figure}

The primary goal of the present paper is to quantify the relative importance of the history effect in a canonical, yet experimentally relevant, configuration that does not exhibit spherical symmetry, namely, that of a single spherical bubble tangent to a horizontal flat plate that grows and dissolves in response to changes in the ambient pressure and in the presence of gravity. In this configuration, the existence of the history effect may become noticeable with a simple experiment: let us consider such a spherical CO$_2$ bubble that dissolves when the pressure is above saturation (see figure \ref{fig:intro_s0}). At a given time $t \approx 60$ s, the pressure is lowered to a new value still above saturation (figure \ref{fig:intro_s0}(b)). Despite the pressure being at all times above saturation, after changing the pressure, the bubble is observed to grow for some time (figure \ref{fig:intro_s0}(a)). Naturally, part of this growth is due to the expansion of the gas. Thus, to observe the effect purely due to diffusion, it is convenient to plot the ambient radius, $R_0$. It is defined as the radius one would observe if the liquid surroundings were at the reference ambient pressure, $P_0$, instead of the actual ambient pressure $P_\infty(t)$:
\begin{equation} \label{eq:Rcorr}
    R_0(t) = R(t) \left(\frac{P_\infty(t)}{P_0}\right)^{1/3}.
\end{equation}
Here, $R(t)$ is the measured bubble radius. Still, the ambient radius can be seen to grow until about $t \approx 100$ s, an effect purely driven by diffusion. Note that $R_0$ was referred to in Part 1 of this article \citep{penas2016} as the pressure-corrected radius $\Rcorr$. However, with the purpose of maintaining the standard nomenclature, $R_0$ will be used throughout this paper. 

This phenomenon may be explained by examining the concentration of dissolved CO$_2$ near the bubble (figure \ref{fig:intro_s0}(c)). Indeed, although the concentration at the bubble surface, given by Henry's law, responds instantaneously to pressure changes, there exists a boundary layer around the bubble where the concentration of CO$_2$ is higher than the instantaneous saturation one, as a result of the dissolution stage that took place before the pressure drop. In the example depicted in figure \ref{fig:intro_s0}(c), it can be seen how the concentration gradient at the bubble's top is actually positive at $t = 65$ s, which explains the growth of the ambient radius. In this figure, numerical simulations like the ones described in \S\ref{sec:formulation}--\S\ref{sec:simulation_results} have been used to compute the concentration field along the $z$ axis. These simulations are validated by comparing the predicted bubble radius with the experimental one (see figure \ref{fig:intro_s0}(a)).

This simple example illustrates that, to properly describe the time evolution of the bubble radius observed in experiments, the history effect must be taken into account. However, a question that was left open in our previous work \citep{penas2016} was the relative importance of this effect in a realistic experimental condition where other effects such as the interference with a wall and natural convection may greatly influence the diffusion-driven bubble dynamics, as was shown by \cite{enriquez2014}. With this idea in mind, another objective of the present work is to propose a numerical approach able to accurately describe the evolution of a bubble attached to a horizontal flat plate and growing/dissolving in the presence of a gravitational field.

While this work only deals with bubbles composed of a single soluble gas, it is important to realise that the history effect is omnipresent in multicomponent bubbles. In Part 1, the history effect was described as `the acknowledgement that at any given time the mass flux across the bubble is conditioned by the preceding time history of the concentration at the bubble interface'. Thus, in dissolving/growing multicomponent bubbles, the flow rate of a particular species across the bubble interface will likely be different from the rest. The species composition inside the bubble will thus change over time, which amounts to time-dependent partial pressures and hence time-dependent interfacial concentrations. It is possible to artificially discern the contribution of the history effect numerically, as was done by \cite{chu2016.1} for the case of a dissolving two-gas bubble. Isolating the history effect experimentally, on the other hand, is anticipated to be much harder.

Finally, it is worth mentioning that the history effect is naturally present in the evaporation of multicomponent drops. \cite{chu2016} have recently developed a formulation that includes a memory integral to describe the diffusion-driven dynamics of multicomponent drops in the presence of a solvent, a phenomenon of relevance in modern techniques of chemical analysis \citep{lohse2016}. In this problem, the faster or slower dissolution of one of the components yields a time-varying composition at the drop's interface, which makes the inclusion of the history integral in Fick's law essential, even when the ambient pressure remains constant.

The paper is structured as follows: \S \ref{sec:experimental} presents the experimental results that illustrate the effect of history in the growth-dissolution of CO$_2$ bubbles tangent to a flat plate. \S \ref{sec:formulation} presents the general mathematical formulation of the problem and sheds light on the importance of the different physical effects involved in the experiments. In \S \ref{sec:numerical_implementation}, a formulation based on the streamfunction--vorticity method is described to simulate the mass transfer and flow field around the bubbles. The simulation results are then presented and discussed in \S \ref{sec:simulation_results}. Finally, \S \ref{sec:conclusions} summarises the main conclusions.

\section{Experimental characterisation of the history effect} 
\label{sec:experimental}
We have carried out experiments to support our theoretical and numerical analyses by subjecting single bubbles to well-controlled, step-like pressure jumps that super- or under-saturate the liquid alternatively. This way, we can make bubbles grow and shrink under repeatable conditions to expose the history effect. It becomes apparent through the differences in the responses to successive identical pressure-time histories.

\subsection{Experimental setup and procedure\label{sec:exp_setup_procedure}}
\begin{figure}
 \centerline{\includegraphics[height=7.5cm]{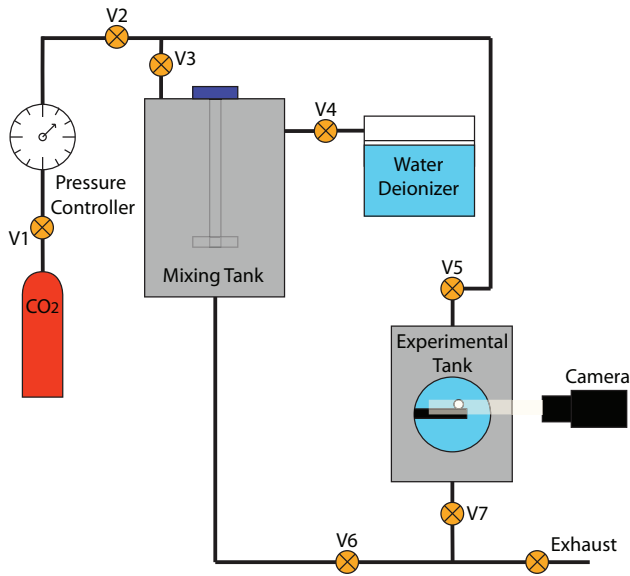}}
  \caption{Sketch of the experimental set-up. See \cite{enriquez2013} for a detailed description\label{setup}.}
\end{figure}

Although the experimental setup has been described in a previous work \citep{enriquez2013}, a brief description is included here for convenience. The facility is fed with water that is demineralised in a purifier (MilliQ A10) and degassed by making it flow through a filter (MiniModule, Liquicel, Membrana). This water enters into the mixing chamber (see figure \ref{setup}), that has been previously flushed with CO$_2$ to purge the air from the system. There the water is stirred in the presence of CO$_2$, kept at the desired saturation pressure, for about 45 minutes. Finally, the experimental tank is pressurised with CO$_2$ at this same pressure and then slowly flooded with the carbonated water, so bubbles do not appear during the filling.
This preparation procedure ensures that in the experimental tank  there are no other gas species present within the liquid or gas phases apart from CO$_2$ (at least in quantifiable amounts).

Placed at the centre of the experimental tank there is a silicon chip, treated to become hydrophilic, with a black-silicon hydrophobic pit (50 $\umu$m in radius) at its centre. The role of this pit is to force a single bubble to nucleate at a fixed location in a repeatable way.
Furthermore, in order to avoid slight temperature variations to affect the diffusion-driven bubble dynamics, the measurement tank is kept at a constant temperature by means of an external chiller.

\begin{figure}
  \centerline{\includegraphics[width=0.9\textwidth]{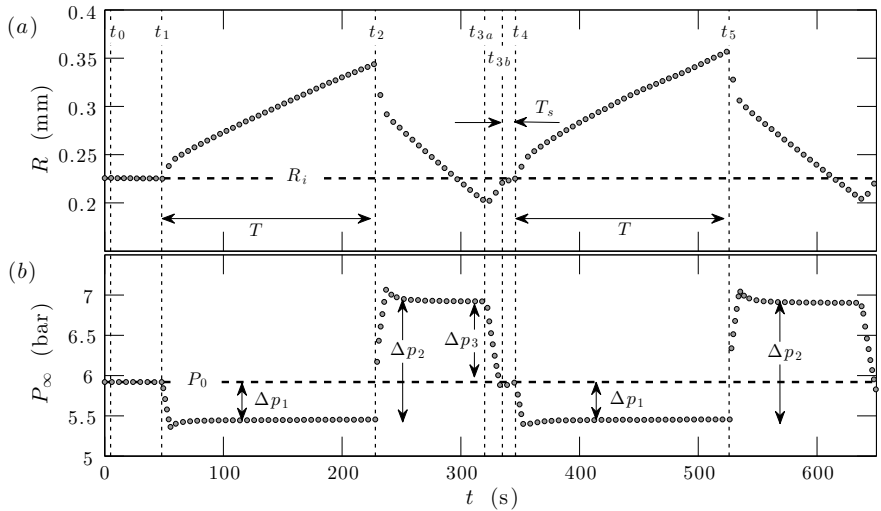}}
  \caption{Experimental procedure during which the bubble is exposed to two identical  supersaturation--undersaturation cycles. The lower plot shows the pressure that the bubble is subjected to, whilst the upper one illustrates its radius time evolution. In few words, once a bubble is stabilised at a given radius $R_i$, it is forced to a growth cycle for a prescribed time $T$ and then to a dissolution cycle down to a size slightly smaller than $R_i$ ($t = t_{3a}$), such that a short time $T_s$ after the pressure returns to the initial level $P_0$ ($t = t_{3b}$), the combination of previous gas expansion (during $t_{3a}$--$t_{3b}$) and history lead the bubble size to the initial radius ($t = t_4$). An identical pressure cycle is immediately imposed, which results in a different time-evolution of the bubble radius due to the history effect.}
\label{fig:procedure}
\end{figure}

Once the measurement tank is filled with the carbonated water, the following experimental procedure is followed:
\begin{enumerate}
\item The pressure is lowered below the saturation value, until a bubble nucleates at the pit and grows up to the desired size, $R_i$.
\item The tank pressure is set again to the saturation value. Then, the pressure is finely adjusted manually until the bubble size does not vary in an observable way for about five minutes. The pressure at which this occurs will be hereafter the one used in the calculations as the saturation pressure. Notice that this procedure allows us to determine the saturation pressure with more accuracy than that given by the pressure controller during the mixing process.\label{ii}
\item At time $t_1$ (see figure \ref{fig:procedure}), the pressure is lowered by a given amount, $\Delta p_1$, during a prescribed time $T = t_2-t_1$. This turns the liquid supersaturated, which leads to bubble growth.\label{iii}
\item Subsequently, at time $t_2$, the pressure is increased by an amount $\Delta p_2$, which causes  undersaturation and the bubble to shrink. \label{iv}
\item When the bubble becomes slightly smaller than $R_i$ at time $t_{3a}$, the pressure is gradually set back to the saturation level $P_0$ ($t_{3b}$) by means of a pressure drop $\Delta p_3$. During this short period ($t_{3a}$--$t_{3b}$) the bubble expands and grows.
\item During a short time $T_s$ after $t_{3b}$, the pressure remains at saturation but the  bubble keeps growing and attains the initial size $R_i$ at $t_4$ due to the history effect (portrayed in figure \ref{fig:intro_s0}). 
At this point ($t_4$), growth step (\ref{iii}) and subsequent dissolution step (\ref{iv}) are immediately repeated at identical $\Delta p_1$ and $\Delta p_2$ respectively.
\end{enumerate}

It is imperative to realise that the pressure and bubble size conditions at $t_4$, just before the pressure jump, are identical to the initial conditions at $t_1$. Namely, $R(t_4) \!=\! R(t_1)$, $P_\infty(t_4) \!=\! P_\infty(t_1)$ and 
$\dd P_\infty(t_4)/\dd t \!=\! \dd P_\infty(t_1)/\dd t \!=\! 0$. The zero-pressure time derivative condition is extremely important to ensure that the bubble is not under the effect of any previous pressure-induced volumetric expansion or compression at the time when $\Delta p_1$ is suddenly imposed. Note that the sole purpose of the pressure change between $t_{3a}$ and $t_{3b}$ and subsequent stabilisation period ($t_{3b}$ and $t_{4}$) is precisely to enforce this last condition. 
This complex procedure allows us to directly isolate and quantify the history effect through direct comparison. Any differences between the first growth rate (during $t_1$--$t_2$) and second growth rate ($t_4$--$t_5$) must be purely attributed to the history effect. The differences arise because at $t_1$ the bubble is in equilibrium with its surroundings (uniform concentration field) and the contribution of history term is essentially negligible. At $t_4$, however, the concentration field surrounding the bubble has evolved. It is no longer uniform, and the bubble is no longer in equilibrium: thus, the contribution of the history term is now larger.

\subsection{Experimental results and discussion}
\begin{figure}
  \centerline{\includegraphics[width=0.9\textwidth]{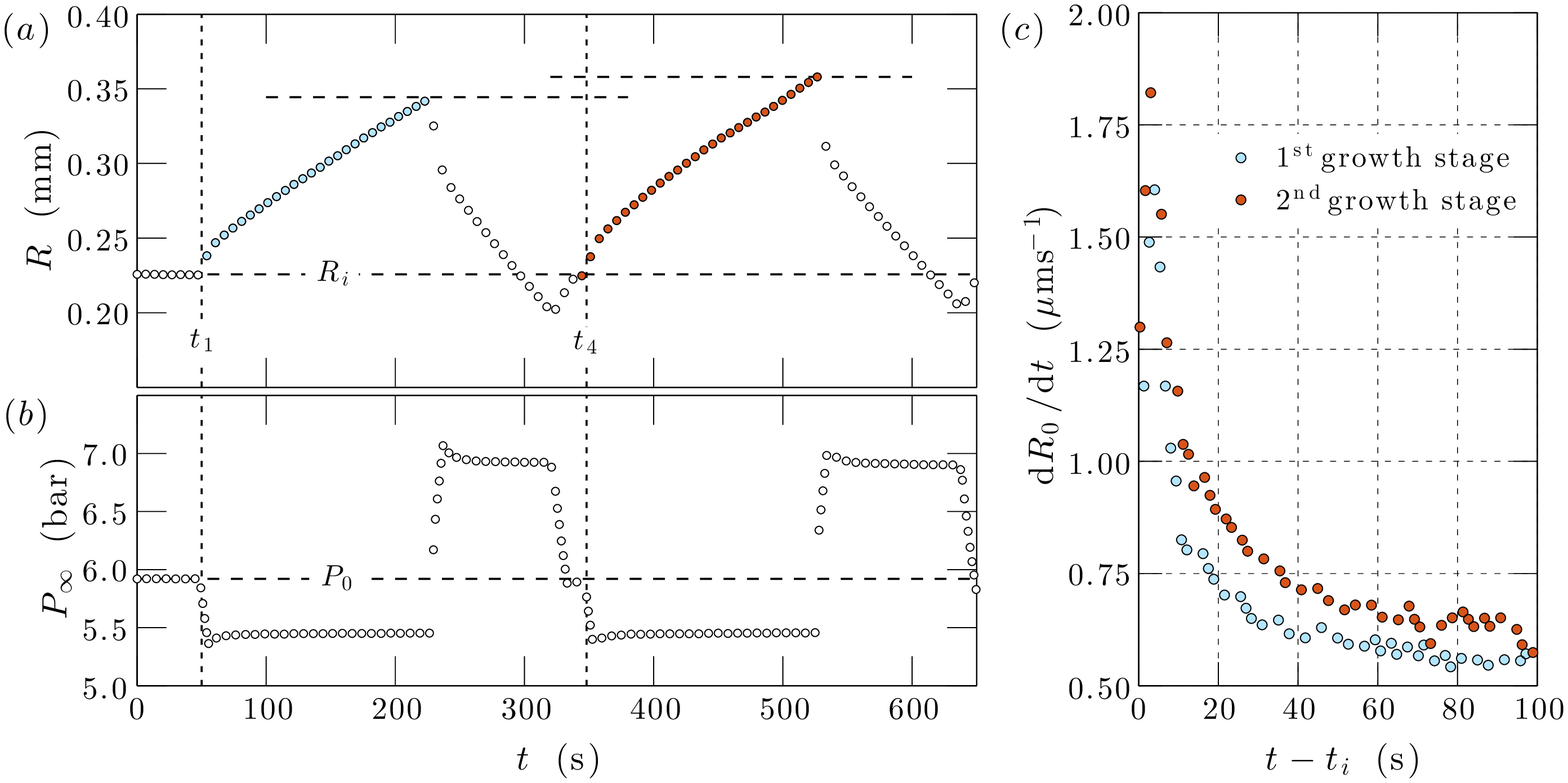}}
  \caption{\label{fig:exps1}Results for experiment 1, showing the time-histories of (a) the measured bubble radius $R$ in response to (b) the imposed pressure $P_\infty(t)$.
In (c), the rate of growth of the ambient radius $R_0$, defined in (\ref{eq:Rcorr}), is plotted for the two growth cycles. The time axis is initialised at $t_1$ or $t_4$ accordingly.}
\end{figure}

\begin{figure}
  \centerline{\includegraphics[width=0.9\textwidth]{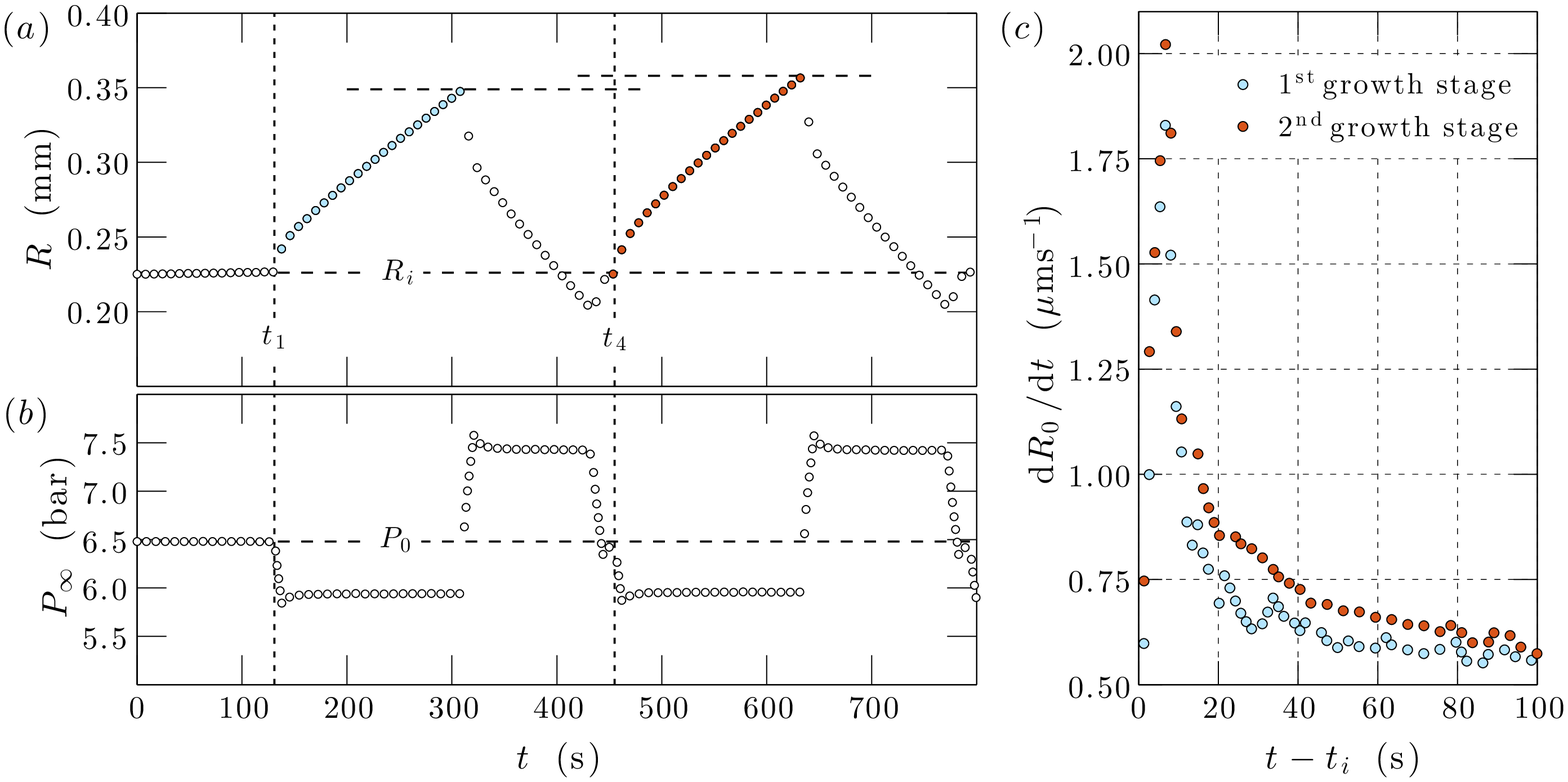}}
  \caption{\label{fig:exps2}Results for experiment 2 (see caption of figure \ref{fig:exps1}). The range of pressures is slightly different to the ones exposed in figure \ref{fig:exps1}. However, the history effect is repeatable. 
  }
\end{figure}

\begin{figure}
  \centerline{\includegraphics[width=0.9\textwidth]{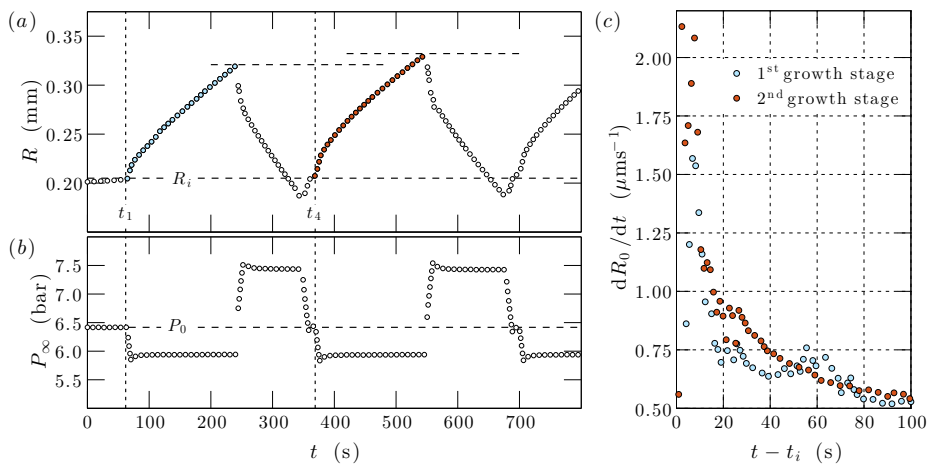}}
  \caption{\label{fig:exps3}Results for experiment 3 (see caption of figure \ref{fig:exps1}). 
  }
\end{figure}

\begin{figure}
  \centerline{\includegraphics[width=0.9\textwidth]{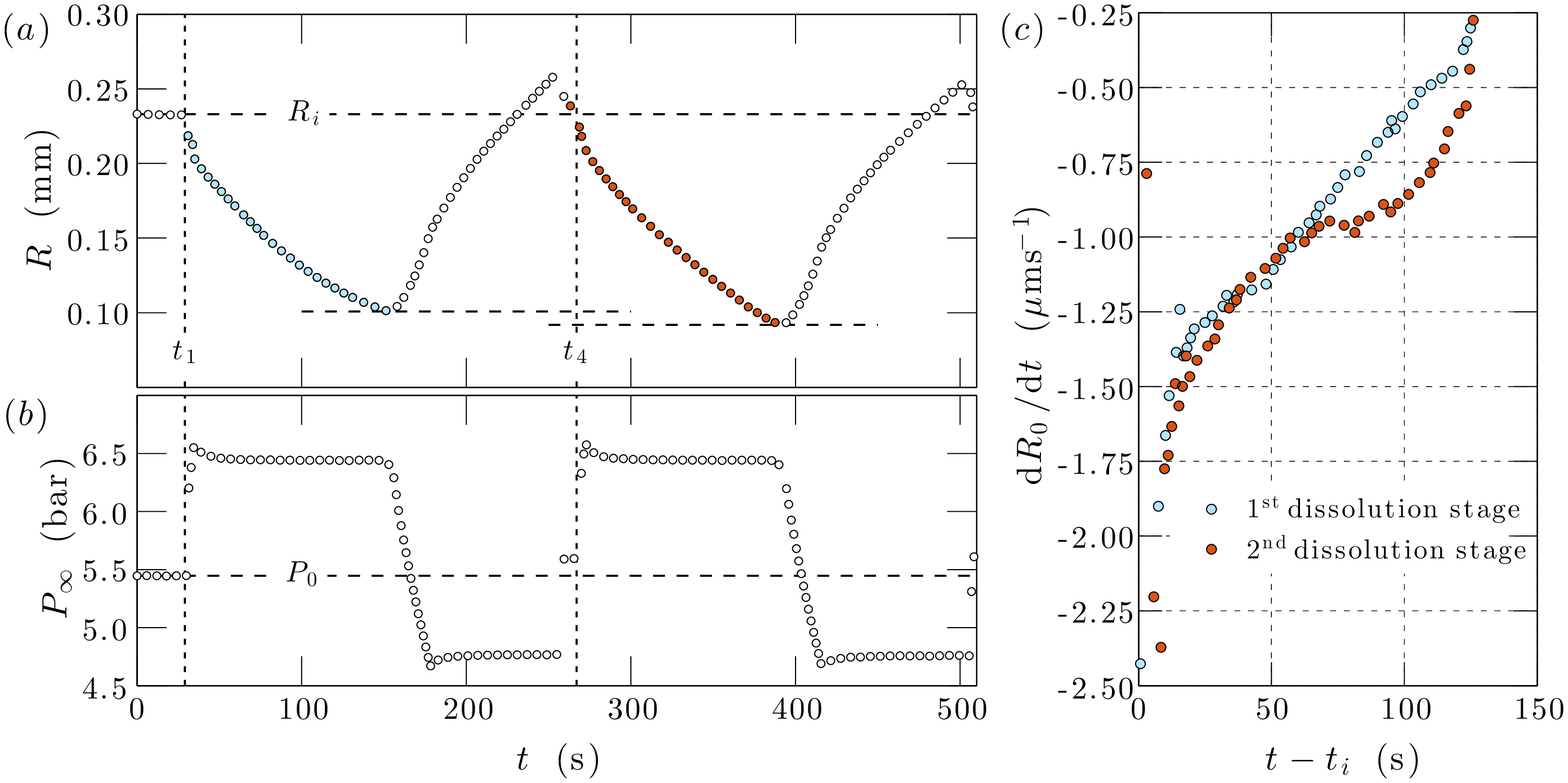}}
  \caption{\label{fig:exps4}Results for experiment 4 (see caption of figure \ref{fig:exps1}). 
  This time, the direction of the pressure jumps is inverted thereby replacing the two growth stages observed in experiments 1--3 with two dissolution stages.}
\end{figure}

In this subsection we present the results of four experiments that manifest the effects of history in bubble growth and shrinkage. Three of the experiments were carried out as described above, while in the fourth the order of the growth and shrinkage stages was swapped, i.e. first shrink and then growing.

In the experiments where the bubble is first made to grow (figures \ref{fig:exps1} to \ref{fig:exps3}), the most apparent difference between the two growth stages is the somewhat larger bubble size achieved at the end of the second stage (see panels labelled as $(a)$). This is a consequence of a more important effect, namely the higher growth rate found during the first instants of the second stage, as predicted by the modified Epstein--Plesset equation with history effects provided in Part 1. To illustrate this point, panels $(c)$ show the time derivative of the ambient bubble radius. In all cases the growth rate during the second stage lies above that of the first one, although both curves eventually converge at longer times, when the memory of the previous dissolution stage damps out. As demonstrated in Part 1, the CO$_2$ accumulated around the bubble during a dissolution stage yields a steeper concentration gradient at the interface that, in turn, leads to a faster growth rate at short times once the pressure is reduced and the liquid is supersaturated again. As the growth progresses, the influence of the initial concentration profile becomes weaker and both growth rates converge to the same curve.

As demonstrated in Part 1, the CO$_2$ accumulated around the bubble during a dissolution stage yields a steeper concentration gradient at the interface that, in turn, leads to a faster growth rate after a short short transient time once the pressure is dropped and the liquid is supersaturated again. As the growth progresses, the influence of the initial concentration profile becomes weaker and both growth rates converge to the same curve. During the very early times after the pressure drop (up to approximately  ten seconds later), the contribution of the history effect on mass transfer is masked by the large growth rates induced by the sudden decrease of the interfacial concentration (induced by this pressure drop via Henry's law) that leads to a steep interfacial concentration gradient. The change in growth rates between the first and second cycles is experimentally indiscernible. This does however agree with theory, as one may observe from the numerically-computed rates in figure 3 of Part 1 \citep{penas2016}.

It is interesting to compare this behaviour with that found when the bubble is forced to first dissolve and then to grow (figure \ref{fig:exps4}). Although unavoidable experimental limitations of the control of the pressure in the facility in this case result in a somewhat noisier time derivative of the ambient radius, the same qualitative behaviour is found. Namely, the magnitude of the rate of change of the radius is larger in the second dissolution stage, thus leading to a smaller radius at the end of this stage. Analogously to what occurred in experiments 1--3, this is a consequence of the local depletion of CO$_2$ near the bubble caused by the intermediate growth stage.

Besides illustrating the history effect in the growth and dissolution of bubbles, these experiments will serve as benchmark cases for the numerical simulations described in the following sections. These numerical analyses will allow us to quantify the relative importance of the different physical effects that play a role in the processes illustrated in figures \ref{fig:exps1} to \ref{fig:exps4} which, besides diffusion, include surface tension, boundary-induced advection and natural convection.

\begin{figure}
  \centerline{\includegraphics[width=0.68\textwidth]{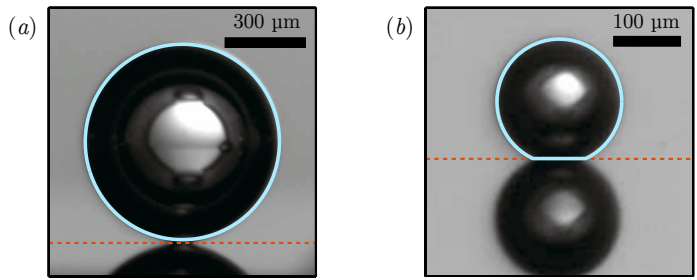}}
  \caption{Bubble snapshots at both extremes of the bubble size range measured during our experiments. The largest radius is (a) $R = 358 \: \umu$m, corresponding to the maximum radius attained during experiment 1 (see later figure \ref{fig:exps1}), whereas (b) $R = 92 \: \umu$m is the smallest radius, obtained during the dissolution experiment 4 (see figure \ref{fig:exps4}). The radius is computed by means of the light-blue circumference fitted to the bubble contour. The horizontal red line marks the height of the bubble-substrate contact line, below which there is the reflection of the bubble on the substrate surface.}
\label{fig:snapshots}
\end{figure}

In the theory that follows, we will assume that the bubble remains strictly spherical at all times. Two experimental snapshots depicting the upper and lower extremes in bubble size are provided in figure \ref{fig:snapshots}. The bubble is actually attached to a cylindrical pit of 50 $\umu$m diameter and 30 $\umu$m depth. The gas volume contained inside the pit can be neglected compared to the total volume of the gas bubble. In the experiments in which the cycles start with a growth phase, where $R> 200 \ \umu$m, the bubble remains spherical, as observed in figure \ref{fig:snapshots}(a). Only at the smallest radii during the experiments starting with a dissolution phase, we observe a spherical cap, figure \ref{fig:snapshots}(b). However, the assumption of perfectly spherical bubble at all time yields a relative error of less than 3 $\%$ as compared to the actual gas volume of the spherical cap and the pit.  Therefore, the assumption of strictly spherical gas bubble for the analysis is more than justified.

\section{Numerical analysis: problem formulation} 
\label{sec:formulation}
\begin{figure}
  \centerline{\includegraphics[width=0.6\textwidth]{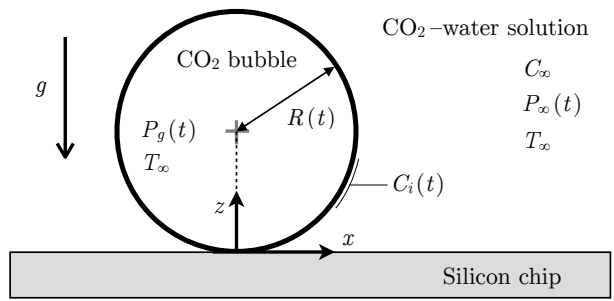}}
  \caption{Sketch of a spherical CO$_2$ bubble adhered to a flat plate. The relevant parameters and functions used in the formulation of the mass transfer problem are also indicated.}
\label{fig:formulation}
\end{figure}

Our goal is to accurately predict the time evolution of the radius of a spherical \cdg bubble adhered to a horizontal flat plate in a CO$_2$-water solution under the action of gravity and variable ambient pressure, as sketched in figure \ref{fig:formulation}. In this section we formulate the mass transfer problem, which involves a non-stationary boundary and that must be coupled with the equations of motion for the liquid assuming axisymmetry around the vertical axis.

\subsection{Mass transfer problem}
The transport of dissolved gas species in the liquid is governed by the following mass transport equation, usually referred to as the advection-diffusion equation,
\begin{equation} \label{eq:mass_transport}
\frac{\partial C}{\partial t}  + \bU \bcdot \bnabla C= D_m \: \nabla^2 C,
\end{equation}
where $C(\bx, t)$ is the molar concentration field, $\bU(\bx, t)$ is the velocity vector field, and $D_m$ is the mass diffusion coefficient. The initial concentration of dissolved gas is assumed to be uniform throughout the liquid and equal to $C_\infty$, equal to the gas concentration in the far-field. The boundary condition of zero-mass flux holds across the impermeable wall.
The concentration boundary condition at the bubble surface, $C_i(t)$, is given by Henry's Law,
\begin{equation}
    C_i (t) = k_H P_g (t),
\end{equation}
where $k_H$ is Henry's (molar based) solubility constant and $P_g(t)$ is the total gas pressure inside the bubble. 
A constant temperature environment $T_\infty$ is assumed, i.e. $k_H$ remains constant, while the ambient pressure $P_\infty (t)$ is set to vary with time $t$.
The bubble gas volume is related to the gas content and pressure via the equation of state for an ideal gas,
\begin{equation} \label{eq:igl}
\frac{4}{3}\pi R^3 P_g = nR_uT_\infty,
\end{equation}
where $n(t)$ is the number of gas moles inside the bubble and $R_u$ denotes the universal gas constant.
The total gas pressure inside the bubble, $P_g$, considering liquid-gas surface tension $\gamma_{lg}$, but neglecting inertial and viscous effects inside the gas phase, is given by 
\begin{equation}
    P_g = P_\infty + 2\gamma_{lg}/R. 
\end{equation}
The mass transfer problem is closed with Fick's first law, which sets the molar flow rate of gas across the bubble surface $S$ to be
\begin{equation} \label{eq:Ficks}
\dot{n} = D\int_S \bnabla C {\boldsymbol{{} \cdot\hat n}}\:\dd  S,
\end{equation}
where $\dd S$ is an infinitesimal area element of the bubble surface, and $\boldsymbol{\hat{n}}$ is the outward-pointing unit normal from the bubble surface.
 
Equations (\ref{eq:mass_transport})--(\ref{eq:Ficks}) represent the mass transfer problem equations.  These must be coupled with the equations of motion from which the velocity field $\bU(\bx,t)$ may be computed.

\subsection{Modelling density-driven natural convection}
The dissolved gas concentration profile around the bubble implies a non-uniform density field of the surrounding liquid-gas solution which may trigger the onset of `density-induced natural convection' \citep{takemura1996}. The change in solution density may be quantified through the concentration expansion coefficient, $\lambda$, usually defined as 
\citep{bataller2009}
\begin{equation} \label{eq:lambda}
\lambda = \frac{1}{\rho_l} \frac{\partial \rho}{\partial C},
\end{equation}
where $\rho_l$ is the density of the pure solvent.
Any change in the solution density is therefore assumed proportional to the change in dissolved gas concentration.
For dilute, mono-solute solutions,  the concentration expansion coefficient is approximately given by (see Appendix \ref{sec:density_change})
\begin{equation} \label{eq:expansion_coeff}
    \lambda \approx \frac{M_g}{\rho_l}- \Vc,
\end{equation}
where $M_g$ is the gas molar mass and $\Vc$ is the (temperature dependent) partial molar volume of the solute in the solvent at infinite dilution.
For \cdg gas in pure water, $\Vc \approx 34.2 \ \mathrm{cm^3/mol}$ \citep{harvey2005}, which results in $\lambda \approx 9.8 \ \mathrm{cm^3/mol}$.

The variations in density considered here are small, of the order of 0.1\%. However, these variations are sufficiently large to have a non-negligible effect on the motion of the flow. Consequently, it was deemed appropriate to take this effect into account via the Boussinesq approximation. This essentially results in the inclusion of a non-uniform buoyancy term imposed by the local dissolved gas concentration into the Navier-Stokes equation (\ref{eq:ns}).
The Boussinesq approximation allows for the flow to be regarded as incompressible when treating the continuity equation.  Therefore, the incompressibility condition,
\begin{equation}
    \bnabla\cdot \bU = 0,
\end{equation}
always holds.
Moreover, the density $\rho(\bx, t)$ of the liquid-gas solution shall be approximated as constant, $\rho(\bx, t) = \rho_l$, in both the inertial and viscous terms of the momentum equation. However, we shall allow small variations in density in the body force (gravity) term. The density field $\rho(\bx, t)$ may then be expressed as
\begin{equation} \label{eq:rho_decomp}
    \rho(\bx, t) = \rho_l + \rho^*(\bx, t)
\end{equation}
where $\rho^*(\bx, t)$ is the density perturbation field arising from the non-uniform concentration field and, evidently, $|\rho^*| \ll \rho_l$.
Similarly, the pressure field in the solution may be decomposed into
\begin{equation} \label{eq:p_decomp}
    P(\bx, t) = P_\infty(t)+ P_h(\bx) + P^*(\bx, t).
\end{equation}
Here $P_h$ is the  background hydrostatic pressure, $\bnabla P_h = \rho_l \bg$, where $\bg$ denotes the gravitational acceleration, and $P^*(\bx, t)$ is the pressure perturbation arising from the fluid motion. It likewise follows that for our experiments, $P_h \ll P_\infty$ and $|P^*| \ll P_\infty$.

\subsection{Equations of motion in terms of the streamfunction and vorticity}
Making use of (\ref{eq:rho_decomp}) and (\ref{eq:p_decomp}), the Navier-Stokes momentum equation may be written as
\begin{equation} \label{eq:ns}
\frac{\partial \bU}{\partial t} +  (\bU \bcdot \bnabla)\bU = 
-\frac{\bnabla P^*}{\rho_l} + \nu \bnabla^2 \bU + \frac{\rho^* \bg}{\rho_l},
\end{equation}
where $\nu$ is the kinematic viscosity of the liquid.
Since the flow is axisymmetric around the vertical ($z$) axis, if we are able to define an orthogonal set of coordinates  $\eta, \xi, \phi$ where $\phi$ is the angle of rotation around the vertical axis, then the velocity field has only two components, $\bU = U_\eta(\eta, \xi) \:\eeta +  U_\xi(\eta, \xi)  \:\exi$, and the whole problem may be treated as two-dimensional.
The vorticity field $\bOmega$ is then also unidirectional and a vorticity scalar, $\Omega$, exists:
\begin{equation} \label{eq:vort_def}
    \bOmega = \Omega \:\ephi  = \bnabla \times \boldsymbol{U}.
\end{equation}
Taking the curl of (\ref{eq:ns}) eliminates the pressure term and the vorticity scalar transport equation is obtained:
\begin{equation}
\frac{\partial \Omega}{\partial t} + \bU \bcdot \bnabla\Omega = \Omega \:\ephi\cdot\bnabla\boldsymbol{U} - 
\nu \bnabla \times \left( \bnabla \times (\Omega \ephi) \right) \bcdot \ephi + \frac{1}{\rho_l} (\bnabla \rho^* \times \bg)\bcdot \ephi.
\end{equation}
It follows from Eq. (\ref{eq:lambda}) that $\bnabla \rho^* = \lambda \rho_l \bnabla C$, so the vorticity transport equation becomes 
\begin{equation} \label{eq:vort_transport}
\frac{\partial \Omega}{\partial t} + h_\phi \, \bU \bcdot \bnabla \left(\Omega/h_\phi\right) = {\cal L}^2\Omega + \lambda (\bnabla C \times \bg)\bcdot \ephi.
\end{equation}
Here, we have made use of the following linear operator:
\begin{equation}
{\cal L}^2 = \frac{1}{h^2}\left[\frac{\partial}{\partial\xi}\left(\frac{1}{h_\phi}\frac{\partial}{\partial\xi}\left(h_\phi\Omega\right)\right) + \frac{\partial}{\partial\eta}\left(\frac{1}{h_\phi}\frac{\partial}{\partial\eta}\left(h_\phi\Omega\right)\right)\right],
\end{equation}
where $h_\phi$ denotes the scale factor in the $\ephi$ direction and $h$ the scale factor in both the $\exi$ and $\eeta$ directions. The coordinate system and scale factors will be introduced quantitatively in \S \ref{sec:numerical_implementation}.
Incompressibility, along with the axisymmetric nature of the flow, allow for the velocity field to be expressed in terms of a scalar stream function, $\Psi$:
\begin{equation} \label{eq:sf_def}
\boldsymbol{U} = \bnabla \times (\Psi/h_\phi\ephi).
\end{equation}
Combining (\ref{eq:vort_def}) and (\ref{eq:sf_def}) results in the following equation for the streamfunction,
\begin{equation} \label{sec:poisson_sf}
{\cal L}^2 \left(\Psi/h_\phi \right) = -\Omega.
\end{equation}
In  \S \ref{sec:numerical_implementation}, it will be shown how the fluid motion may be obtained by simultaneously solving $\Psi$ from (\ref{sec:poisson_sf}) and $\Omega$ from (\ref{eq:vort_transport}) numerically by employing a stream function-vorticity method in dynamic tangent-sphere coordinates.
It will be seen that the boundary conditions for $\Psi$ and $\Omega$  can be determined from those for $\bU$ through careful analysis. From the physical point of view, the velocity field must satisfy the kinematic and dynamic (zero-shear stress) boundary conditions along the moving bubble boundary, in addition to the no-slip condition at the wall.

\subsection{On the parameters and timescales of the problem}
\label{sec:timescales}
This subsection intends to shed light on the physics governing the diffusion-driven growth and dissolution of a bubble attached to a flat plate. More specifically, the goal is to prove that the concentration and velocity fields evolve over very disparate timescales, which will allow for an efficient procedure to numerically solve the problem formulated in previous subsections.

The processes involved in this problem introduce four characteristic timescales: $t_s$ for bubble growth and dissolution, $t_m$ for mass diffusion of the dissolved gas, $t_v$ for viscous diffusion of momentum and $t_b$ for the density-induced convection. Let $U$ denote the characteristic flow velocity. When the advection induced by the moving boundary dominates over natural convection, then $U$ is the interface velocity $U_s \sim \dot R$. When convection overcomes boundary-induced advection, then $U$ becomes the convection velocity $U_b$.
The characteristic lengthscale is the bubble radius $R$. 
For mass-diffusion-controlled growth driven by a molar concentration difference $\Delta C$ between the bubble boundary and the bulk fluid, the flow behaviour may be characterised using three dimensionless parameters. These are the Jakob  \citep{szekely1971} and Grashof numbers for mass transfer, in addition to the Schmidt number, defined as follows:
\refstepcounter{equation}
$$ 
\Ja = \frac{M_g |\Delta C|}{\rho_g}, \qquad
\Gr  =  \frac{\gra R^3}{\nu^2}, \qquad
\Sc  =  \frac{\nu}{D_m},
\eqno{(\theequation{\mathit{a-}\mathit{c}})}
$$
where $\rho_g$ is the density of the gas bubble and $g$ is the magnitude of the acceleration due to gravity. \Ja may be regarded as a measure of the driving force for bubble growth induced by the concentration difference and gas solubility. \Gr represents the ratio of buoyancy (convection) and viscous forces. \Sc is the ratio of momentum and mass diffusivities.

Here we shall consider bubble growth or dissolution that is primarily driven by mass diffusion. We may then use the approximate result obtained by \cite{epstein1950} or \cite{scriven1959} to estimate bubble growth as
\begin{equation} 
R \sim \Ja \sqrt{D_{m}t}.
\end{equation}
It then follows that the bubble growth timescale and boundary-induced advection velocity scales are
\refstepcounter{equation}
$$ 
U_s   = \dot R \sim \frac{\Ja^2 D_m}{R}, \qquad
t_s   = \frac{R}{U_s} \sim \frac{R^2}{\Ja^2 D_m}.
\eqno{(\theequation{\mathit{a},\mathit{b}})}
$$
The magnitudes of the terms in the mass transport equation (\ref{eq:mass_transport}) are 
\refstepcounter{equation}
$$
\frac{\partial C}{\partial t} \sim \frac{|\Delta C|}{t_m}, \quad
\bU \bcdot \bnabla C \sim \frac{U |\Delta C|}{R}, \quad
D\nabla^2 C \sim \frac{D_m |\Delta C|}{R^2}
\eqno{(\theequation{\mathit{a-}\mathit{c}})} \label{eq:mass_mag}
$$
where $t_m$ is the characteristic time required for a significant concentration change over characteristic lengthscale $R$. Similarly, taking $\Omega \sim U/R$, the magnitudes of the terms in the vorticity transport equation (\ref{eq:vort_transport}) are
\refstepcounter{equation}
$$
\frac{\partial \Omega}{\partial t} \sim \frac{U}{R \: t_v}, \quad
\bU \bcdot \bnabla \Omega \sim \frac{U^2}{R^2}, \quad
\nu{\cal L}^2 \Omega \sim \frac{\nu U}{R^3}, \quad \lambda \bnabla C \times \bg \sim \frac{\gra}{R},
\eqno{(\theequation{\mathit{a-}\mathit{d}})} \label{eq:vort_mag}
$$
where similarly $t_v$ refers to the time required for a significant vorticity change over the same characteristic lengthscale $R$. For $Sc\sim 1$ or $Sc\gg1$, the characteristic convection velocity and timescale may be obtained from a balance between the viscous term (\ref{eq:vort_mag}{\it{c}}) and the buoyancy term (\ref{eq:vort_mag}{\it{d}}) in the vorticity transport equation,
\refstepcounter{equation}
$$ 
U_b \sim \frac{\gra R^2}{\nu}, \qquad
t_b = \frac{R}{U_b} \sim \frac{\nu}{\gra R}.
\eqno{(\theequation{\mathit{a},\mathit{b}})}
$$
The ratio of velocities is given by $U_b/U_s = \Gr \:\Sc /\Ja^2$. The ratio of the advection term (\ref{eq:mass_mag}{\it{b}}) and the diffusive term (\ref{eq:mass_mag}{\it{c}}) in the mass transport equation yields a P\'eclet number, $\Pe = UR/D$.
The ratio of the advection term (\ref{eq:vort_mag}{\it{b}}) over the diffusive term (\ref{eq:vort_mag}{\it{c}}) in the vorticity transport equation similarly yields a Reynolds number, $\Rey = UR/\nu$.
Neglecting natural convection, setting $U = U_s$ gives $\Pe = \Ja^2$ and $\Rey = \Ja^2/\Sc$.
Likewise, natural convection dominating over boundary-induced advection, $U = U_b$, results in $\Pe = \Gr \: \Sc$ and $\Rey = \Gr$.

From the above analysis, we may conclude that mass and momentum diffusion will clearly dominate over advection and natural convection provided $\Gr \: \Sc < 1$ and $\Ja < 1$ (i.e.  $\Pe$ and $\Rey$ are small). In such a case, the mass diffusion and viscous timescales are the leading timescales in the mass transport and vorticity transport equations respectively. The unsteady term in each transport equation may then be balanced by the corresponding diffusive term, yielding
\refstepcounter{equation}
$$
t_m \sim \frac{R^2}{D_m}, \qquad t_v \sim \frac{R^2}{\nu}.
\eqno{(\theequation{\mathit{a},\mathit{b}})} \label{eq:tm,tv}
$$
The ratio between the mass diffusion timescale and the other timescales are
\refstepcounter{equation}
$$ 
\frac{t_m}{t_s} = \Ja^2, \qquad
\frac{t_m}{t_b} = \Gr \: \Sc, \qquad
\frac{t_m}{t_v} = \Sc.
\eqno{(\theequation{\mathit{a-}\mathit{c}})}\label{eq:timescale_ratios}
$$
As reference for the conditions explored in this work, a \cdg gas bubble with $R = 0.25$ mm  growing in a 15\%  supersaturated CO$_2$-water solution at 5 bar and 293 K, with $\lambda = 9.8 \ \mathrm{cm^3/mol}$, results in
$\Ja = 0.12$, $\Gr = 0.038$ and  $Sc = 523$. The Rayleigh number is $\Gr\: \Sc = 19.6$.
Under these conditions, intentionally similar to those of our experiments, (\ref{eq:timescale_ratios}) translates to
\begin{equation}
t_s > t_m \sim t_b \gg t_v.
\end{equation}

The vorticity/velocity field around a bubble evolves at a timescale $t_v$ provided by the viscous diffusion of momentum. This timescale is much faster than the timescale $t_m$ of mass transfer, i.e. the time required to observe a significant change in the concentration field surrounding the bubble. Likewise, the timescale $t_s$ in which a substantial change in the bubble radius may be observed is significantly larger than $t_m$. This means that the thin boundary layer approximation (valid when $t_s \ll t_m$, i.e. $Ja \gg 1$), while suitable for treating the fast growth of bubbles in highly supersaturated liquids \citep{rosner1972}, is clearly not applicable here.

The timescale of interest is of course $t_m$. Let us neglect density-driven convection for the moment. At every timestep of this slow timescale $t_m$, provided $t_m \gg t_v$ ($Sc \gg 1$), viscous action ensures that the flow always reaches (over a much faster timescale $t_v$) a steady-state solution. In other words, at every timestep of $t_m$, the advection term in the mass transport equation may then be computed from the steady-state vorticity (hence velocity) solution imposed by the instantaneous concentration field and interface velocity. We shall refer to it as the quasi-steady advection approximation.
It is worth pointing out that bubbles of other gases with solubility parameter, $\Lambda$, smaller than that of CO$_2$ (such as Nitrogen or Oxygen) can be described as well with this approximation, as $t_m$ will be even much smaller than $t_s$. In these cases, the history effect --which is a diffusive effect-- will be even more apparent since boundary-driven advection will have a smaller influence.

Considering now density-driven convection, provided $t_b \gg t_v$ ($\Gr \ll 1$), then viscosity is able to establish a quasi-steady velocity field in a time much shorter than that taken by buoyancy to induce changes in the flow. In other words, although buoyancy must be taken into account to properly compute the velocity field around the bubble, it does not affect the validity of the quasi-steady advection approximation. This knowledge will now be used in the next section when implementing the equations into a numerical model.

\section{Numerical analysis: implementation} 
\label{sec:numerical_implementation}
\subsection{Non-dimensionalisation}
We begin by introducing the dimensionless time, radius and Cartesian coordinates, ambient pressure and mole number:\refstepcounter{equation}
$$ 
    \tau = \frac{D_m}{R_i^2}t, \qquad 
    a = \frac{R}{R_i}, \qquad
    \tilde{\bx} = \frac{\bx}{R_i}, \qquad
    p = \frac{P_\infty}{P_0}, \qquad
    \mu = \frac{R_u T_\infty}{4/3 \pi R_i^3 P_0} n.
\eqno{(\theequation{{\mathit{a-}\mathit{d}}})}
$$
In this work we have chosen the characteristic radius $R_i$ to be the initial radius $R(t=0)$. Similarly, the characteristic ambient pressure $P_0$ corresponds to the initial liquid pressure, $P_\infty(0)$, whereas the mole number $R_u T_\infty n$ is made dimensionless with that contained in a bubble of radius $R_i$, immersed in a liquid at pressure $P_0$ and in the absence of surface tension.
Note that the timescale of choice has been that of mass diffusion, $t_m$, presented in (\ref{eq:tm,tv}{\it a}). Additionally, the molar concentration field $C$ and the interfacial molar concentration $C_{i}$ may be nondimensionalised through
\refstepcounter{equation}
$$ 
    c = \frac{C-C_\infty}{k_H P_0}, \qquad 
    c_i = \frac{C_i-C_\infty}{k_H P_0}.
\eqno{(\theequation{\mathit{a},\mathit{b}})}
$$
The dimensionless counterparts of the vorticity scalar, velocity and streamfunction are
\refstepcounter{equation}
$$ 
\omega = \frac{R^2_i}{D_m}\Omega, \qquad
\bu = \frac{R_i}{D_m} \bU, \qquad 
\psi = \frac{1}{R_i D_m} \Psi. 
\eqno{(\theequation{\mathit{a-} \mathit{c}})}
$$
Lastly, let us present the following dimensionless parameters and dimensionless numbers:
\refstepcounter{equation}
$$ 
    \Upsilon = \frac{C_\infty}{k_H P_0}, \qquad
    \Lambda = k_H R_u T_\infty , \qquad
    \sigma = \frac{2\gamma_{lg}}{R_i P_0}, \qquad
    \Gr_0 = \frac{\lambda k_H P_0 g R_i^3 }{\nu^2}.
\eqno{(\theequation{\mathit{a-} \mathit{d}})}
$$
The parameter $\Upsilon$ refers to the initial saturation level of the solution, $\Lambda$ is a solubility parameter, $\sigma$ is the surface tension parameter, while $\Gr_0$ is a reference Grashof number, $\Gr_0 = \Gr\:k_H P_0/|\Delta C|$. 

\subsection{The tangent-sphere coordinate system}
\label{sec:coord_system}
\begin{figure}
  \centerline{\includegraphics[width=0.6\textwidth]{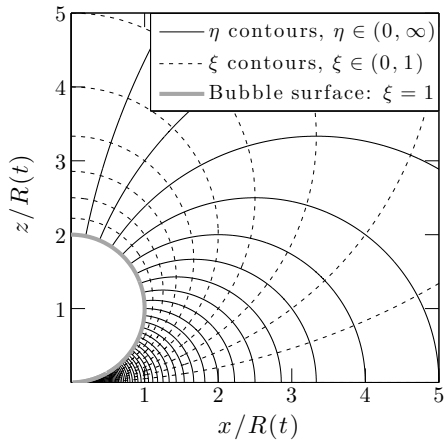}}
  \caption{Contourlines of the tangent-sphere $\eta$, $\xi$ coordinates, plotted in the $y=0$ ($\phi=0$) Cartesian plane. $\eta = 0$ lies on the $z$-axis, $\eta \rightarrow \infty$ at the contact point. The horizontal wall lies on the $\xi = 0$ isosurface, while the bubble surface is always mapped by $\xi = 1$. The separation of the plotted contours is uniform ($\Delta\eta = \Delta \xi = 0.1$).}
\label{fig:ts_coordinates}
\end{figure}

The problem can be conveniently recast in dimensionless tangent-sphere spatial coordinates $(\eta, \xi, \phi)$, where
\refstepcounter{equation}
$$ 
    \xad = 2a\frac{ \eta}{\eta^2+\xi^2} \cos\phi, \qquad 
    \yad = 2a\frac{\eta}{\eta^2+\xi^2} \sin\phi, \qquad
    \zad = 2a\frac{\xi}{\eta^2+\xi^2}.
\eqno{(\theequation{\mathit{a-} \mathit{c}})}
$$
The contours of $\eta$ and $\xi$ satisfy the following inverse relations \citep{moon1988, batchelor1979}:
\refstepcounter{equation}
$$ 
   \xad^2 + \yad^2+\zad^2 = (2 a/\eta)\sqrt{\xad^2 + \yad^2}, \quad 
   \xad^2 + \yad^2+\zad^2 = \left(2a/\xi\right)\zad.
   \eqno{(\theequation{\mathit{a}, \mathit{b}})} \label{eq:contours}
$$
These coordinates, represented in figure \ref{fig:ts_coordinates}, scale with the dimensionless radius of the bubble, $a(\tau)$. The scale factors are defined as 
\refstepcounter{equation}
$$ 
    \hti = \frac{h_\eta}{R_i} = \frac{h_\xi}{R_i} = \frac{2 a}{\eta^2+\xi^2}, \qquad 
    \displaystyle \htiphi = \frac{h_\phi}{R_i} = \frac{2a\eta}{\eta^2+\xi^2}. 
    \eqno{(\theequation{\mathit{a}, \mathit{b}})} \label{eq:scale_factors}
$$

The partial time derivative of any scalar function $f$ described by fixed Cartesian coordinates $(x,y,z)$ expands as the material derivative when described by our $R(t)$-scaling spatial coordinates ($\eta, \xi$). Taking the partial derivative of $f$ with respect to time $\tau$, we find
\begin{equation} \label{eq:matder}
    \frac{\partial}{\partial \tau} f(\xad,\yad,\zad,\tau) 
    = \frac{\dD}{\dD \tau}f(\eta(\tau),\xi(\tau), \tau) 
    = \frac{\partial f}{\partial \tau} 
    +\eta'\frac{\partial f}{\partial \eta} 
    +\xi'\frac{\partial f}{\partial \xi}.
\end{equation}
The prime notation ($'$) denotes $\dd/\dd\tau$. The terms containing $\eta'$ and $\xi'$ represent the apparent advection of a quiescent fluid relative to our time-varying coordinate system. 
Let us define the \textit{a priori} unknown corresponding (dimensionless) apparent velocity field as $\bu_{rel}(\eta, \xi, \tau)= u_{rel, \eta}\:\eeta + u_{rel, \xi}\:\exi$.
The advection term on $f$ would then be
\begin{equation} \label{eq:reladv}
    (\bu_{rel} \bcdot \btinabla)f = 
    \frac{u_{rel,\eta}}{\hti}\frac{\partial f}{\partial \eta} 
    +\frac{u_{rel,\xi}}{\hti}\frac{\partial f}{\partial \xi}, 
\end{equation}
where the operator $\btinabla = R_i \bnabla$ is dimensionless. 
Comparing (\ref{eq:matder}) and (\ref{eq:reladv}) immediately reveals that $u_{rel,\eta} =  \hti \eta'$ and $u_{rel,\xi}=  \hti \xi'$.
Thus, the dimensionless velocity field of our scaling coordinate system (relative to any fixed point in the physical domain) is just equal to 
$-\bu_{rel} = -(\hti\eta' \: \eeta + \hti\xi' \: \exi)$.
Differentiating (\ref{eq:contours}{\it a}) and (\ref{eq:contours}{\it b}) independently with respect to $\tau$, one finds that
\refstepcounter{equation}
$$ 
\eta' = a'\eta/a, \qquad \xi' = a'\xi/a.
    \eqno{(\theequation{\mathit{a}, \mathit{b}})} \label{eq:eta'_xi'}
$$

\subsection{Streamfunction--vorticity formulation}
\label{sec:psi-omega_formulation}
The streamfunction satisfies the following equation,
\begin{equation}
    \tilde{\cal L}^2 (\psi/\htiphi) = -\omega,
\end{equation}
which makes use of the dimensionless operator $\tilde{\cal L}^2 = R_0^2 {\cal L}^2$, and that may be rewritten in terms of the coordinates $\eta$ and $\xi$ as
\begin{equation} \label{eq:streamfunction_ts}
\frac{\eta^2-\xi^2}{2 a \htiphi^2}\frac{\partial \psi}{\partial \eta} 
+ \frac{\eta \xi}{a \htiphi^2} \frac{\partial \psi}{\partial \xi} 
+ \frac{1}{\hti^2 \htiphi} \left(\frac{\partial^2 \psi}{\partial \eta^2} +\frac{\partial^2 \psi}{\partial \xi^2}\right) 
= -\omega.
\end{equation}
Boundary conditions are derived and explained in Appendix \ref{sec:bcs}. These are
\refstepcounter{equation}
$$ 
    \psi(\eta,0) = 0, \quad
    \psi(\eta,1) = -\frac{2 a^2 a'}{\eta^2+1}, \quad
    \psi(0,\xi) =  -2 a^2 a',\quad
    \psi(\infty,\xi) = 0.
\eqno{(\theequation{\mathit{a-} \mathit{d}})}
$$
Once $\psi(\eta, \xi)$ is known, from the definition in (\ref{eq:streamfunction_ts}), the dimensionless velocity components may be computed by numerically differentiating $\psi$ according to 
\refstepcounter{equation}
$$ 
\gdef\thesubequation{\theequation a,b} \label{eq:u_psi}
u_\eta = \frac{1}{\htiphi \hti} \frac{\partial \psi}{\partial \xi}, \quad
u_\xi = -\frac{1}{\htiphi \hti} \frac{\partial \psi}{\partial \eta}.
    \eqno{(\theequation{\mathit{a}, \mathit{b}})} 
$$

Expressions for the velocity field in the Cartesian reference frame are included in Appendix \ref{sec:transformation_matrix}. The dimensionless vorticity transport equation reads
\begin{equation}
    \frac{\partial \omega}{\partial \tau} + \htiphi\,\bu \bcdot \btinabla \left(\frac{\omega}{\htiphi}\right) = 
\Sc\:\tilde{\cal L}^2 \omega + \Gr_0 \:\Sc^{2}\: (\btinabla c \times \ghat) \bcdot \ephi,
\end{equation}
where $\ghat = \bg/g$. The Rayleigh number $\Gr_0\:\Sc^{2}$ represents the ratio between the reference buoyant transport of momentum and diffusive transport of mass. In ($\eta$, $\xi$) coordinates, the vorticity transport equation becomes 
\begin{subequations} \label{eq:vort_transport_ts}
\begin{equation}
\frac{\partial \omega}{\partial \tau} = 
    L \frac{\partial \omega}{\partial \eta} 
    + M \frac{\partial \omega}{\partial \xi} 
    + P \left(\frac{\partial^2 \omega}{\partial \eta^2} +\frac{\partial^2 \omega}{\partial \xi^2}\right)
    + Q + S \omega,
\end{equation}
\end{subequations}
\addtocounter{equation}{-1}
where $L$, $M$, $P$, $Q$ and $S$ are time and space-dependent coefficients given by  
\refstepcounter{equation}
$$ 
    L = - \Sc \: \frac{\eta^2-\xi^2}{2 a \htiphi} -\frac{a'\eta}{a} - \frac{u_\eta}{\hti}, \qquad
    M = - \Sc \: \frac{\xi\eta}{a\htiphi} -\frac{a'\xi}{a} - \frac{u_\xi}{\hti},
\eqno{(\theequation{\mathit{b}, \mathit{c}})}
$$
\addtocounter{equation}{-1}
\refstepcounter{equation}
$$ 
    P =  \frac{\Sc}{\hti^2}, \qquad
    Q = - \Gr_0 \: \Sc^2 \: \left(\frac{\eta^2-\xi^2}{2a} \frac{\partial c}{\partial \eta} 
    + \frac{\eta\xi}{a}\frac{\partial c}{\partial \xi}\right),
\eqno{(\theequation{\mathit{d}, \mathit{e}})}
$$
\addtocounter{equation}{-1}
\refstepcounter{equation}
$$ 
    S =  -\frac{\Sc}{\hti}\,\frac{\eta^2+2\xi^2+\xi^4/\eta^2}{\left(\eta^2+\xi^2\right)^2} - \frac{1}{\hti\eta}\,\frac{u_\eta\left(\eta^2-\xi^2\right) + 2u_\xi\,\eta\xi}{\eta^2+\xi^2}.
\eqno{(\theequation{\mathit{f}})}
$$
Coefficient $Q$ contains the density-driven convection term. The cross product was evaluated by first expressing $\ghat$ in its ($\eeta$, $\exi$) components via the transformation matrix described in Appendix \ref{sec:transformation_matrix}.
Boundary conditions are derived and explained in Appendix \ref{sec:bc_vorticity}. These are
\refstepcounter{equation}
$$ 
    \frac{\partial \omega}{\partial \xi}(\eta,0) = \frac{\hti u_n}{Sc \:\Delta\tau_v}, \qquad
    \frac{\partial \omega}{\partial \eta}(\eta,1) = \frac{4a'\eta}{a(\eta^2+1)} + \frac{2}{a}u_\eta(\eta,1),
\eqno{(\theequation{\mathit{a}, \mathit{b}})} \label{eq:bc_vorticity}
$$
\addtocounter{equation}{-1}
\refstepcounter{equation}
$$ 
    \omega(0,\xi) =  0,\qquad
    \frac{\partial \omega}{\partial \eta}(\infty,\xi) = 0.
\eqno{(\theequation{\mathit{c}, \mathit{d}})}
$$
Here, $\Delta\tau_v$ is the computational timestep for the viscous transport of momentum. We shall report its meaning in \S \ref{sec:num_procedure}.

\subsection{Formulation for the mass transfer problem}
In dimensionless form, the advection-diffusion equation (\ref{eq:mass_transport}) becomes 
\begin{equation}
\frac{\partial c}{\partial \tau} + \bu \bcdot \btinabla c = \tinabla^2 c,
\end{equation}
or equivalently,
\begin{subequations} \label{eq:mass_transport_ts}
\begin{equation}
\frac{\partial c}{\partial \tau} = 
    F \frac{\partial c}{\partial \eta} 
    + G \frac{\partial c}{\partial \xi} 
    + H \left(\frac{\partial^2 c}{\partial \eta^2} +\frac{\partial^2 c}{\partial \xi^2}\right),
\end{equation}
\end{subequations}
\addtocounter{equation}{-1}
with
\refstepcounter{equation}
$$ 
    F = - \frac{\eta^2-\xi^2}{2 a \hti \eta} -\frac{a'\eta}{a} - \frac{u_\eta}{\hti}, \qquad
    G = - \frac{\xi}{a\hti} -\frac{a'\xi}{a} - \frac{u_\xi}{\hti}, \qquad
    H = \frac{1}{\hti^2}.
\eqno{(\theequation{\mathit{b-} \mathit{d}})}
$$
The velocity field components $u_\eta(\eta,\xi, \tau)$ and $u_\xi(\eta,\xi, \tau)$ must of course be computed beforehand as detailed in section \ref{sec:psi-omega_formulation}. Boundary conditions for the concentration are
\refstepcounter{equation}
$$ 
    \frac{\partial c}{\partial \xi}(\eta,0) = 0, \quad
    c(\eta,1) = c_i, \quad
    \frac{\partial c}{\partial \eta}(0,\xi) = 0, \quad
    \frac{\partial c}{\partial \eta}(\infty,\xi) = 0. \quad
\eqno{(\theequation{\mathit{a-} \mathit{d}})} \label{eq:bc_concentration}
$$
The interfacial concentration $c_i(\tau)$, appealing to Henry's Law, is given by
\begin{equation} \label{eq:c_s}
 c_i = \left(p + \frac{\sigma}{a}\right)-\Upsilon .
\end{equation}
The dimensionless molar flow rate $\mu'$ across the interface may be computed non-dimensionalising (\ref{eq:Ficks}), resulting in
\begin{equation} \label{eq:mu'}
\mu' = -3 \Lambda a  
    \int_0^\infty \frac{\eta}{1+\eta^2} 
    \left.\frac{\partial c}{\partial \xi} \right|_{\xi = 1} \dd \eta.
\end{equation}
Finally, the last equation remaining is the dimensionless ideal gas law, 
\begin{equation} \label{eq:igld}
    \left( p + \frac{\sigma}{a}\right)a^3 = \mu.
\end{equation}

\subsection{Numerical procedure}
\label{sec:num_procedure}
The experiments were simulated by numerically solving the governing equations presented in the previous section. To do so, we used a second order finite-difference discretisation in space and an implicit Euler method in time. The latter was chosen in search for unconditional stability, bearing in mind that the grid spacing becomes infinitesimally small as $\eta \rightarrow \infty$.

We have seen in \S \ref{sec:timescales} that this problem involves multiple scales. The governing equations have been non-dimensionalised in time with the timescale $t_m = R_i^2/D_m$ for mass diffusion, $\tau = t/t_m$. The mass transport equation will therefore require a timestep $\Delta\tau_m$ that suitably advances within $t_m$. 
However, the momentum transport equation requires a much smaller timescale $t_v$ and consequently, $\Delta\tau_v \sim  \Delta\tau_m/\Sc$. Stability of the scheme requires to advance in $\Delta\tau_v$. Doing so, however, the overall number of time iterations and computational cost required to span the whole duration of the experiments would be exceedingly high.

This issue could be solved making use of the quasi-steady advection approximation, validated in \ref{sec:timescales}. Essentially, this approximation allows to advance the simulation in $\Delta\tau_m$ rather than in $\Delta\tau_v$. 
After timestep $\Delta\tau_m$, time advances from $\tau_n$ to $\tau_{n+1}$, where subscript $n$ refers to the $n$-th time iteration. Given the actual concentration field $c_n$, suppose mass transfer across the bubble interface results in a change of radius from $a_n$ to $a_{n+1}$ with corresponding rate $a'$. What is the vorticity field $\omega_{n+1}$ for this new configuration? 
We first make the initial guess: $\omega_{n+1} = \omega_n$. The vorticity field is then allowed to independently evolve through the vorticity transport equation, advancing with timestep $\Delta\tau_v$. By virtue of the quasi-steady advection approximation, the concentration field, bubble radius and velocity are treated as invariants. After $k$ iterations, the vorticity converges to the steady-state solution. 
We used the following criterion for convergence:
\begin{equation} \label{eq:convergence}
\varepsilon^2_{max} = \mathrm{max}\left\{\frac{(\omega_{k+1}-\omega_k)^2}{\mathrm{max}(\omega_k^2)} \right\} < 10^{-6}.
\end{equation}
The resulting vorticity and velocities are then used to solve the advection-diffusion equation to find $c_{n+1}$. The process is then repeated for the following timestep. 
This way, the overall number of iterations and computational time are greatly reduced.

The numerical procedure followed at every time iteration $n$ consists of the following steps:
\begin{enumerate}
    \item{ Update time $\tau_n$ and primary variables: vorticity $\omega_n$, concentration $c_n$ and mole number $\mu_n$. Define the timestep $\Delta\tau_m$.}
    \item{ Obtain value of the externally applied pressures $p_n$ and $p_{n+1}$ through linear interpolation of the experimental pressure data.}
     \item{ (\ref{eq:igld}) yields a cubic equation, $p_n a_n^3 + \sigma a_n^2 - \mu_n = 0$. Solve to obtain the radius $a_n$.}
     \item{ Compute $\mu'_n$ from (\ref{eq:mu'}) and integrate over $\Delta\tau_m$ to obtain $\mu_{n+1}$.}
     \item{ Obtain the radius $a_{n+1}$ in the same way as in step (iii) and then compute $a'$.}
     \item{ Compute $\omega_{n+1}$ and velocity components $u_\xi$ and $u_\eta$ with the streamfunction--vorticity method. First set $\omega_{k=0} = \omega_n$. Define the secondary timestep $\Delta\tau_v$. 
     \begin{enumerate}
     \item{Update $\omega_k$.}
         \item{ Compute $\psi$ from (\ref{eq:streamfunction_ts}) using $\omega_k$.}
         \item{Compute $u_\xi$ and $u_\eta$ from (\ref{eq:u_psi}).}
         \item{Compute $\omega_{k+1}$ from (\ref{eq:vort_transport_ts}).}
         \item{Check for convergence} of $\omega_{k+1}$.  If tolerance in (\ref{eq:convergence}) is met, $\omega_{n+1} = \omega_{k+1}$; otherwise, update $k$ and return to step ({\it a}).
     \end{enumerate}}
     \item{Compute $c_{n+1}$ from (\ref{eq:mass_transport_ts}).}
     \item{ Update $n$ and return to step (i).}    
\end{enumerate} 

\begin{table}
  \begin{center}
\def~{\hphantom{0}}
\begin{tabular}{l l c l l l l l l l}    
                                                                   &\qquad \qquad &Exp.   &$R_i$    &$P_0$     &$T_\infty$    &$\Upsilon$    &$\Lambda$    &$\sigma\!\times\!10^{3}$    &$\Gr_0$\\    
Physical properties                                                &              &       &mm       &bar       &$^\circ$C     &              &             &                            &   \\[3pt]
$k_H = 3.36 \times 10^{-4} \ \mathrm{mol \: m^{-3} \: Pa^{-1}}$    &              &1      &0.226    &5.92      &21.6          &1.0           &0.823        &1.05                        &0.218    \\
$D_m = 1.92 \times 10^{-9} \ \mathrm{m^2 \: s^{-1}}$               &              &2      &0.226    &6.48      &21.8          &1.001         &0.824        &0.96                        &0.240    \\            
$\gamma_{lg} = 7.0\times 10^{-2} \ \mathrm{N \: m^{-1}}$           &              &3      &0.205    &6.42      &22.6          &1.004         &0.826        &1.06                        &0.177    \\
$\lambda = 9.8\times 10^{-6} \ \mathrm{m^3 \:mol^{-1}}$            &              &4      &0.233    &5.45      &21.9          &1.0           &0.824        &1.10                        &0.221    \\
$\nu = 1.004 \times 10^{-6} \ \mathrm{m^2 \:s^{-1}}$               &              &       &         &          &              &              &             &                            &    \\
    \end{tabular}
    \caption{Values of the parameters used in the simulations, corresponding to the experiments discussed in \S \ref{sec:experimental}. For completeness, the Schmidt number is $\Sc \!=\! \nu/D_m \!=\!  523$.}
    \label{tab:sim_parameters}
  \end{center}
\end{table}

\section{Simulation results and discussion}
\label{sec:simulation_results}
\begin{figure}
  \centerline{\includegraphics[width=\textwidth]{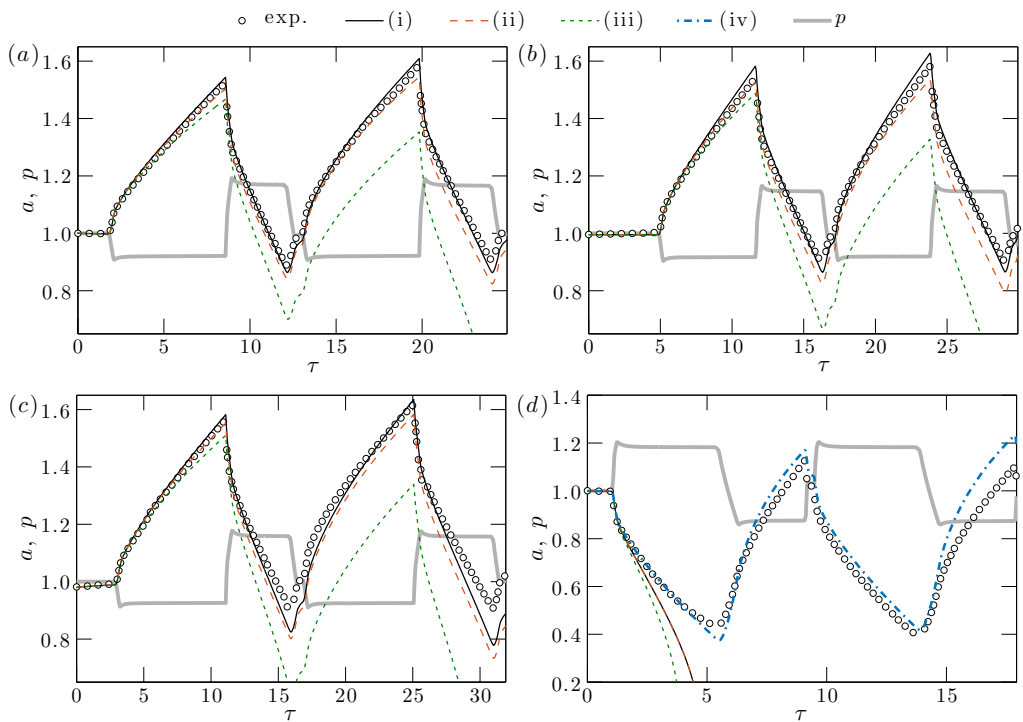}}
  \caption{Time evolution of the dimensionless bubble radius, $a(\tau)$. Simulation (black curves) is compared to experimental data (markers) for experiments (a) 1, (b) 2, (c) 3 and (d) 4. The time evolution of the dimensionless ambient pressure, $p(\tau)$, is also included (grey curve). The different simulation curves correspond to (i, iv) the full solution, (ii) solution where density-induced convection is neglected, (iii) solution for pure diffusion (both density-induced convection and boundary-induced advection are neglected). Moreover, case (iv) is only used to model experiment 4 as seen in (d), corresponding to the full solution coupled with the \cdg stratification model, with $\Upsilon_\mathit{wall} = 1.25$.}
\label{fig:expsim_comp}
\end{figure}

The simulation predictions for the bubble size history are compared with the experiments in figure \ref{fig:expsim_comp}. The simulation input parameters  consist of the physical properties for \cdg gas and water, together with specific reference parameters for each experiment. These are listed in table \ref{tab:sim_parameters}. The saturation level of the far-field, $\Upsilon = C_\infty/k_HP_0$, was accurately determined from the initial evolution of the radius time-history of each experiment before the first jump in pressure, as described in \S\ref{sec:experimental}.

Case (i) in figure \ref{fig:expsim_comp} corresponds to the full solution. For cases (ii) and (iii), density-induced convection is neglected: $\lambda = 0$, thus $Gr_0 = 0$. This translates to setting $Q = 0$ in the vorticity transport equation (\ref{eq:vort_transport_ts}). Additionally (iii) is the solution for pure diffusion, i.e. without any velocity. This implies setting $u_\eta = u_\xi= a' = 0$ in the mass transport equation (\ref{eq:mass_transport_ts}). Consequently, solution (iii) does not make use of the streamfunction--vorticity formulation and the mass transfer problem can be solved independently. Examination of panels (a), (b) and (c) (corresponding to experiments 1--3) of  reveals that taking into account the interface-induced advection is essential in order to reproduce the experimental results beyond the first growth stage. Note that this holds though the P\'eclet numbers here are small, in fact of the order of the dimensionless pressure jumps, i.e. around $0.1-0.2$. It can be concluded that, although the instantaneous rate of mass transfer may only be slightly affected by advection, its effect accumulates over time and becomes important to describe the evolution of the bubble when subjected to successive expansion--compression cycles. In fact, neglecting advection will always yield smaller bubble sizes: during growth, advection squeezes the concentration boundary layer around the bubble, thus increasing the concentration gradient and therefore the mass transfer rate of gas into the bubble. Analogously, during shrinkage, advection stretches the boundary layer and smoothens up the gradient, which results in a dissolution slower than that predicted for pure diffusion. In consequence, advection effects accumulate during both growth and dissolution, which makes them noticeable over long times. Contrarily, our numerical results show that taking into account convection barely modifies the calculated radius time evolution. We will get back to the effect of convection below.

Experiment 4, whose results are shown in figure \ref{fig:expsim_comp}(d), deserves special attention. The simulation predicts a complete dissolution of the bubble after the first jump in pressure. Interestingly, the experimental dissolution rate is much slower. We hypothesise that there exists a thin stably-stratified layer of thickness $\lesssim R_i$ above the substrate oversaturated with \cdg. This high-density layer can easily form during the compression--expansion cycles used to stabilise the bubble size described in \S \ref{sec:exp_setup_procedure}. Indeed, we have observed in Laser-Induced Fluorescence experiments reported elsewhere \citep{PenasLopez_etalPRF2016} that such a layer can form in a matter of seconds during bubble shrinkage. Notice that the accumulation of \cdg near the substrate becomes more effective at inhibiting bubble dissolution as the bubble becomes smaller. Contrarily, for the case of bubble growth, such a layer would affect the mass flux in a region that corresponds to the ``dead zone'' proposed by \cite{enriquez2014} (see discussion below), which would explain why its existence barely affects experiments 1--3. Thus, for experiment 4, the far-field concentration can no longer be taken as uniform. In fact, we may speculate that this layer is characterised by a vertical concentration gradient bounded by $C_\mathit{wall}$ at the wall and $C_\infty$ at the unstratified, bulk fluid.

Our streamfunction--vorticity method has the limitation that the simulation domain covers a small region in the vicinity of the bubble. Imposing a stratified concentration field as the initial condition is ineffective, since the surplus of dissolved CO$_2$ in this small liquid volume is promptly engulfed by the bubble and a uniform concentration field is quickly established.

\begin{figure}
\centerline{\includegraphics[width=0.55\textwidth]{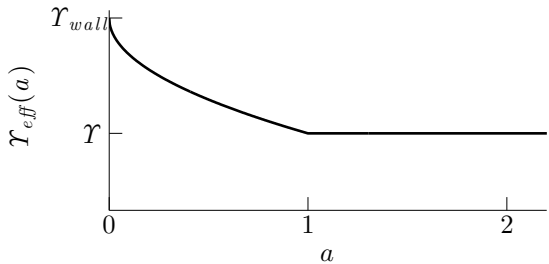}}
  \caption{Sketch of the ``effective" far-field concentration as a function of the instantaneous bubble radius as a means to model the effect of stratification on the mass transfer rate across the bubble interface}
\label{fig:upsiloneff}
\end{figure}

We may bypass this limitation by modelling the effect of stratification essentially through just an effective increase (decrease) of mass transfer towards (from) the bubble. It consists in imposing a reduction on $\Delta C = C_i - C_\infty$ as the bubble becomes sufficiently small. This is done by replacing $C_\infty$ by an ``effective" far-field concentration, $C_{\infty, \mathit{eff}} (R) \geq C_\infty$, that depends on the current size of the bubble. As the bubble shrinks, the effective far-field concentration increases and so $\Delta C_\mathit{eff} = C_i - C_{\infty, \mathit{eff}}$ gets positively closer to zero (slower dissolution) or becomes more negative (faster growth). We propose a concentration profile for the dimensionless ``effective" far-field concentration of the form
\begin{equation}
\Upsilon_\mathit{eff} = \left\{
\begin{array}{lcl}
\Upsilon_\mathit{wall} + (\Upsilon - \Upsilon_\mathit{wall})\: a^{1/2}, & \mathrm{for}  & a<1 \\
\Upsilon , & \mathrm{for}  & a \ge 1
\end{array}
\right.
\end{equation}
where $\Upsilon_\mathit{eff} = C_{\infty, \mathit{eff}}/k_HP_0$ and the free parameter $\Upsilon_\mathit{wall} = C_\mathit{wall}/k_HP_0$ is the extrapolated saturation level at the wall. It is sketched in figure \ref{fig:upsiloneff}. Note that the ``effective" concentration profile proposed does not, by any means, represent the actual stratification profile that one would observe experimentally. The exponent of $1/2$ on $a$ is chosen arbitrarily, on the grounds that the effect of stratification becomes stronger closer to the wall. For our purposes, a linear relation (exponent of 1 on $a$) would nonetheless yield similar results.

The replacement of $\Delta C$ by $\Delta C_\mathit{eff}$ can be easily implemented in the mass transfer problem equations through the concentration boundary condition at the bubble interface (\ref{eq:bc_concentration}{b). The dimensionless interfacial concentration defined in (\ref{eq:c_s}) now becomes
\begin{equation}
    c_\mathit{i,eff} = c_i - (\Upsilon_\mathit{eff} - \Upsilon)
    = p + \sigma/a - \Upsilon_\mathit{eff}.
\end{equation}

For the simulation corresponding to experiment 4, $\Upsilon_\mathit{eff}$ was found to vary within a maximum of 10\% when taking $\Upsilon = 1.0$ and $\Upsilon_\mathit{wall} = 1.25$. Nevertheless, it has a remarkable effect on the mass transfer rates across the bubble in our simulations, as the bubble shrinks to a size comparable to the hypothesised thickness of the layer. We stress that this artificial approach nonetheless portrays the physical significance that a slightly oversaturated layer close the substrate can have a big impact on the bubble dissolution rate.

We focus now on the history effect. As explained in the introduction, this effect is a manifestation of the influence in the instantaneous concentration field of the previous growth or dissolution stages that the bubble has been subjected to. To illustrate the occurrence of this effect in the current configuration, figure \ref{fig:cfields2comp} shows a comparison of the concentration field for experiment 2 (cf. figure \ref{fig:exps2} and later figure \ref{fig:expsim_comp}(b)) obtained at the (a) first and (b) second growth stage at the instant when the bubble radius is of the same size: $a = 1.05$.
The history effect on the growth rate is evident: the concentration contours in (b) are noticeably closer together than those in (a).  This thinner shell translates to steeper gradients and increased mass transfer, which explains the faster growth rates observed in the second cycle in figures \ref{fig:exps1}(c), \ref{fig:exps2}(c) and \ref{fig:exps3}(c).

\begin{figure}
  \centerline{\includegraphics[width=0.75\textwidth]{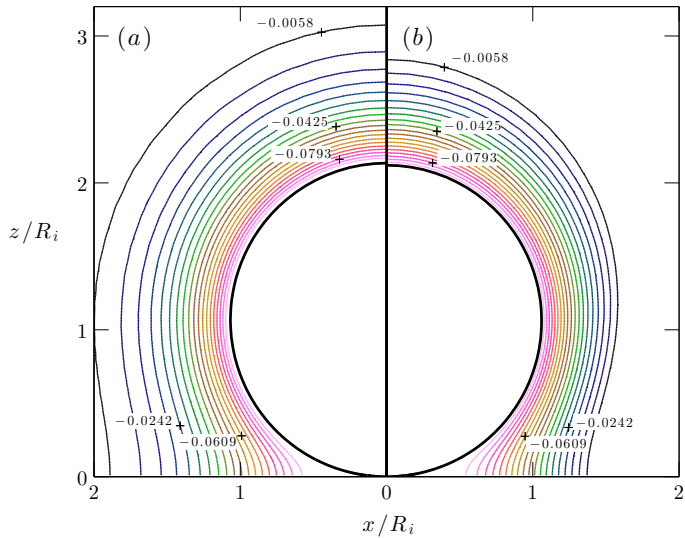}}
  \caption{Dimensionless concentration field contours for experiment 2 according to  simulation at the instant of time when the bubble radius is $a = 1.05$ during the (a) first growth stage and (b) second growth stage. The darker concentration contours (from pink to green to dark blue) indicate positively increasing values.}
\label{fig:cfields2comp}
\end{figure}

The results of our simulations can also be used to validate the hypothesis made by \cite{enriquez2014} about the existence of a ``dead zone'' near the contact point where mass transfer is almost zero. As these authors show, the growth rate of a bubble attached to a plate can be computed by considering that the mass-flux is uniform along the bubble surface, as given by the spherically-symmetric solution, except in a region close to the contact point between the bubble and the plate, where it is nearly zero. The boundary of this region is given by the intersection of the plate with a sphere of radius $R + \sqrt{\pi D_m t}$, which approximately corresponds to the outer limit of the concentration boundary layer around the bubble. Therefore, the overall mass flux across the bubble interface can be modelled as if the mass exchange occurred only across an effective area that excludes this zone. In figure \ref{fig:plateeffect} we plot the local mass-flux distribution along the bubble surface for different time instants during the first growth cycle of experiment 2. The step-like markers indicate the start of the dead zone computed using the model by \cite{enriquez2014}. It can be seen that the effective area predicted by this model agrees fairly well with the region where the mass transfer is nearly uniform, specially at short times.

\begin{figure}
  \centerline{\includegraphics[width=0.65\textwidth]{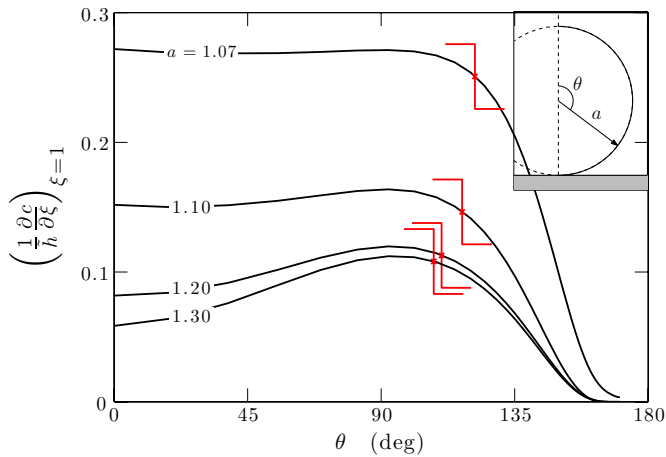}}
  \caption{Dimensionless mass-flux across the bubble interface as a function of the angle $\theta$ for different times during the first growth stage of experiment 2. The step-like markers indicate the angle $\theta^*$ delimiting the effective bubble area available for mass transfer, where $\cos(\theta^*) = -a/(a+\sqrt{\upi\tau})$ according to \cite{enriquez2014}. }
\label{fig:plateeffect}
\end{figure}

Finally, we come back to the role of natural convection. Figure \ref{fig:flow_nb_t1} portrays the structure of the flow and concentration field around a growing bubble assuming there is no natural convection. The boundary layer at the wall due to the no-slip condition is highly distinguishable, as is the vorticity generation at the bubble boundary. It follows that the structure of the flow field is identical for growth and dissolution, except that the direction of the flow velocities are reversed. On the other hand, density-induced natural convection greatly modifies the structure of the flow, as one may observe from figure \ref{fig:flow_b}. Natural convection breaks down the symmetry of the flow structure when comparing growth against dissolution. In dissolution, a low-velocity recirculation region surrounding the bubble is observed. As a result, the concentration field is stretched upwards in growth and compressed downwards in dissolution.
However, despite the changes that convection induces in the velocity field, its effect on the concentration boundary layer near the bubble is minute, as is revealed by the comparison between figures \ref{fig:flow_nb_t1}(c) and \ref{fig:flow_b}(c). Consequently, under the conditions investigated here, its effect on the bubble radius is barely noticeable.
It should be pointed out that, would the growth stage last for longer times, the relatively large value of the Rayleigh number of these experiments ($Ra \sim 20$) suggests that convection should contribute significantly to the growth rate of the bubble, based on recent results by \cite{dietrich2016}. Indeed, in figure 8 of that paper, it can be appreciated how for a Rayleigh number of about 20, the Sherwood number exhibits a noticeable difference with respect to the value for very small Rayleigh. Although the configuration that they explore, a sessile droplet of a liquid heavier than the environment, is different, it is reasonable to conclude that a similar influence might be expected here if the bubble was left to grow indefinitely.
 
\begin{figure}
  \centerline{\includegraphics[width=0.9\textwidth]{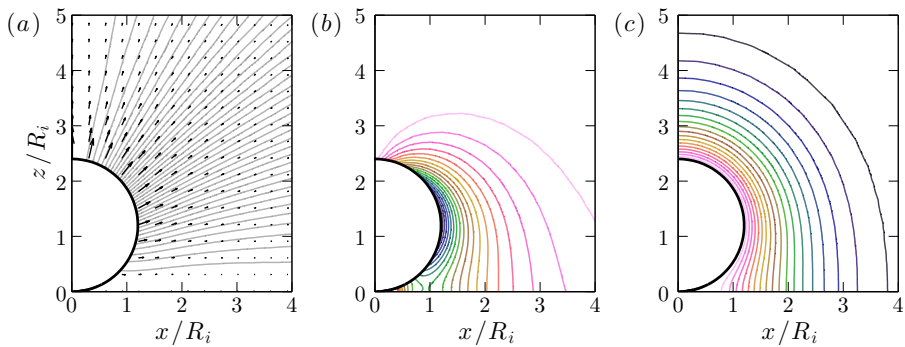}}
  \caption{Simulation snapshots for solution (ii) of experiment 1, in which natural convection has been neglected. The snapshots are taken at $\tau = 3.5$, corresponding to the first growth stage (see figure \ref{fig:expsim_comp}(a)). These show (a) the velocity field (arrows) and streamlines, (b) the vorticity field contours and (c) the concentration field contours. The darker  vorticity and concentration contours (from pink to green to dark blue) indicate positively increasing values.}
\label{fig:flow_nb_t1}
\end{figure}

\begin{figure}
  \centerline{\includegraphics[width=0.9\textwidth]{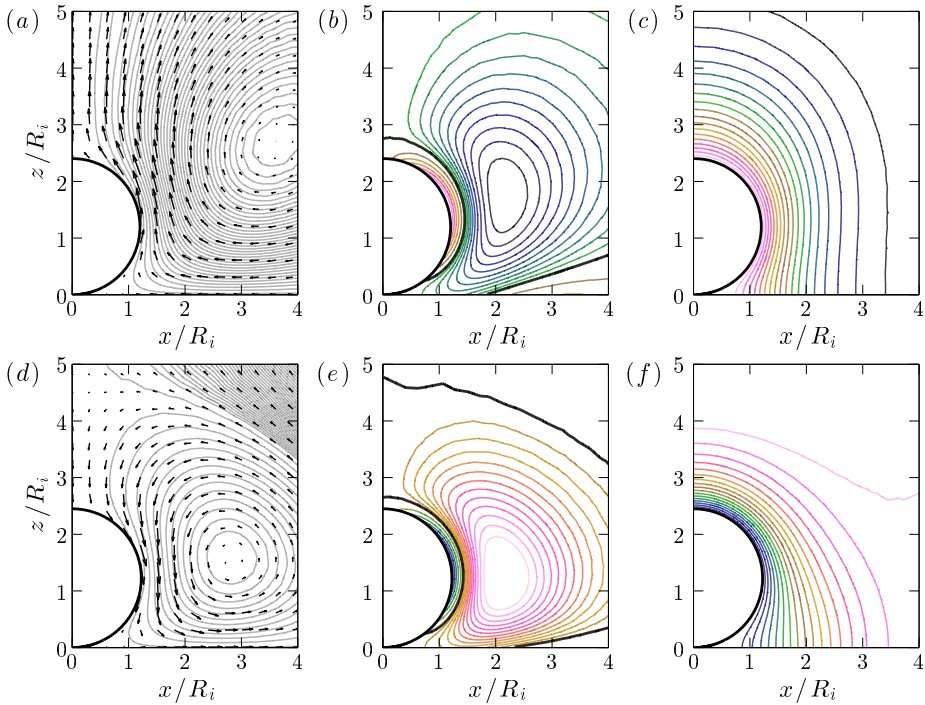}}
  \caption{Simulation snapshots for solution (i) of experiment 1, which takes into account natural convection. The snapshots of (a) the velocity field and streamlines, (b) the vorticity field contours and (c) the concentration field contours are taken at $\tau = 3.5$, corresponding to the first growth stage (see figure \ref{fig:expsim_comp}(a)). Snapshots (d)--(f) show the same fields as (a)--(c) above, but are taken at $\tau = 9.5$, corresponding to the first shrinkage stage. The darker  vorticity and concentration contours (from pink to green to dark blue) indicate positively increasing values. The thick black contourlines in (b) and (e) mark the zero-vorticity contour. Despite the significant changes that natural convection induces in the velocity field, its influence on the concentration field in the vicinity of the bubble is minute, as is revealed by the comparison of the iso-concentration lines with and without convection (panel (c) vs. figure \ref{fig:flow_nb_t1}(c)).}
\label{fig:flow_b}
\end{figure}

\section{Conclusions}
\label{sec:conclusions}
We have experimentally and numerically explored the influence of the past history of the ambient pressure experienced by a bubble on its instantaneous rate of mass transfer---the so-called history effect. This effect is caused by a history-induced pre-existing concentration boundary layer of dissolved gas that surrounds the bubble at the beginning of a given growth or dissolution stage.

To illustrate the existence of the history effect in practical situations, we report here several experimental results. Firstly, we show that the mass of the bubble, represented by the ambient radius $R_0$, can experience transient growths even when the (varying) pressure is kept above saturation at all times. We would naively expect that such a situation would lead to a monotonic dissolution, since the liquid is undersaturated during the whole process.

Secondly, by subjecting the bubble to two consecutive identical expansion--compression cycles, we are able to observe how the history effect becomes manifest in a higher growth rate at the beginning of the second cycle. Thirdly, we report an additional experiment in which the order of the expansion--compression stages has been swapped, obtaining analogous results.

Subsequently, a vorticity--streamfunction formulation has been developed to accurately describe the experimental results reported in the paper, and can be applied to other situations of practical interest in areas such as microfluidics or carbon sequestration. 

By performing order of magnitude analyses, we show that our experiments belong to a regime dominated by mass and viscous diffusion. Moreover, the flow around a growing/dissolving bubble in presence of natural convection can be considered quasi-steady, since the viscous timescale is much faster than the timescale of mass transfer. Thus, the momentum equation can be decoupled from the mass transfer problem. The simulations performed with this strategy are able to describe accurately the experimental results in most cases. 

One of the most important lessons learnt from these simulations is that boundary-induced advection needs to be taken into account if the bubble radius is to be described accurately. The reason for this is that advection enhances growth and diminishes dissolution, thus its effects accumulate to yield larger bubble sizes. Regarding natural convection, we have seen that it greatly modifies the overall structure of the flow around the bubble, albeit its influence on the concentration boundary layer near the bubble surface---where mass transfer takes place---is only subtle under the conditions explored here, i.e. a bubble subjected to successive growth--dissolution periods rather than one that is left to growth for long times.

Finally, we must point out that our simulation strategy does not describe well the case of a bubble that first dissolves and then grows. An explanation is proposed to fix this problem, which consists in assuming the existence of a stably-stratified, CO$_2$-rich fluid layer that accumulates on top of the plate. This way, as the bubble size becomes of the order of the layer thickness, the apparent bulk concentration that the bubble sees is higher. Therefore, it dissolves more slowly than theory predicts. Our estimations suggest that even slight excesses in the apparent bulk concentration may have a strong effect in the dissolution rate of the bubble.\\

The authors acknowledge the support of the Spanish Ministry of Economy and Competitiveness through grants DPI2014-59292-C3-1-P and DPI2015-71901-REDT, partly financed by European funds. This work was also supported by the Netherlands Centre for Multiscale Catalytic Energy Conversion (MCEC), an NWO Gravitation programme funded by the Ministry of Education, Culture and Science of the government of the Netherlands. 

\appendix
\section{On the density change with concentration}
\label{sec:density_change}
For dilute solutions far from the solvent's critical point, the partial molar volume of the solvent can be taken as its molar volume in the pure state, and the solute partial molar volumes  are independent of concentration \citep{harvey2005}. In addition, the concentration of the solute is then approximately proportional to its mole fraction $x_g$, where $x_g \ll 1$ ($x_g = 0.003$ for a CO$_2$--water solution saturated at 5 bar).

Let us then consider a binary solution at pressure $P_0$ consisting of a single solute $(\cdg)$ and solvent (pure water). 
At the experimental conditions, the ionisation of \cdaq into \hco and H$^+$ ions may be neglected since ionic concentrations are low. In fact, we may prove this by defining a molar dissociation ratio (where we first make the dilute solution approximation that molality is proportional to molarity) as
\begin{equation}
    \frac{x_\mathrm{HCO_3^-}}{x_\mathrm{CO_2}} = \frac{x_\mathrm{H^+}}{x_\mathrm{CO_2}} \approx \sqrt{\frac{K}{x_\mathrm{CO_2}}}
    = \sqrt{\frac{K}{k_HV_lP_0}},
\end{equation}
where $K$ is the equilibrium constant ($K = 4.17 \cdot 10^{-7}$ at 20$^\circ$C), $k_H$ is Henry's solubility constant and $V_l$ is the partial molar volume of pure water. 
We see that the dissociation ratio decreases with pressure or total \cdg concentration. For pressures from 1 to 8 bars, the ionised form only accounts for around 2.5$\%$ down to 1\%  of the total \cdg in solution, respectively. 

Considering thus a non-ionic binary liquid-gas solution, Henry's Law relates the molar concentration and the mole fraction of the solute to its partial pressure $P_0$ by
\refstepcounter{equation}
$$ 
    C = k_H P_0, \qquad x_g = k_H V_l P_0,
\eqno{(\theequation{\mathit{a}, \mathit{b}})}
$$
where $V_l$ denotes the partial volume of the solvent.
The solution density may be obtained from 
\begin{equation}
    \rho = \rho_l+\Delta\rho  = \frac{x_l M_l + x_g M_g}{x_l V_l + x_g \Vc}, \quad
    \mbox{with} \quad x_g + x_l = 1.
\end{equation}
Here $M_l$ and $M_g$ denote the molar masses of solvent and solute, while $\Vc(T)$ is the partial molar volume of the gas at infinite dilution.
For dilute solutions, i.e. in the limit $x_g \rightarrow 0$, the density will change with concentration according to
\begin{equation}
\frac{\partial \rho}{\partial C} \approx \lim_{x_g \rightarrow 0} \: V_l\frac{\partial \rho}{\partial x_g} 
= \lim_{x_g \rightarrow 0} \: V_l \frac{V_lM_g-M_l\Vc}{[V_l + (\Vc - V_l)x_g]^2}
= M_g - \frac{M_l\Vc}{V_l}.
\end{equation}
The expression in (\ref{eq:expansion_coeff}) for the concentration expansion coefficient immediately follows.

\section{Boundary conditions}
\label{sec:bcs}
\subsection{Boundary conditions for velocity}
\label{sec:bc_velocity}
The velocity field at the moving bubble boundary must satisfy both the kinematic and dynamic boundary conditions. The kinematic boundary condition refers to the continuity of the velocity component normal to the interface. 
We recall that the bubble boundary is described by $\xi = 1$ at all times. The interface normal velocity is thus $u_\xi(\eta, 1)$ and it must be exactly equal to $-u_{rel, \xi}(\eta,1)$ as derived in \S \ref{sec:coord_system}. Therefore,
\begin{equation} \label{eq:bc_kinematic}
    u_\xi(\eta,1) = -\hti \xi'.
\end{equation}
The dynamic boundary condition refers to the continuity of tangential stress across the interface. Neglecting the viscosity of the gas, this condition reads
\begin{equation} \label{eq:bc_dynamic}
    \frac{\partial}{\partial \eta}\left( \frac{u_\xi}{\hti} \right) +
    \frac{\partial}{\partial \xi}\left( \frac{u_\eta}{\hti} \right) = 0 \quad \mbox{on} \quad \xi = 1.
\end{equation}
The remaining set of boundary conditions for the dimensionless velocity field are no-slip at wall, zero flow velocity at the contact point and symmetry conditions along the vertical ($z$) axis:
\refstepcounter{equation}
$$ 
u_\eta(\eta, 0) = 0, \quad u_\xi(\eta, 0) = 0, \qquad u_\eta(\infty, \xi) = 0, \quad u_\xi(\infty, \xi) = 0
\eqno{(\theequation{\mathit{a-} \mathit{d}})}
$$
\addtocounter{equation}{-1}
\refstepcounter{equation}
$$ 
u_\eta(0, \xi) = 0, \quad \frac{\partial u_\xi}{\partial \eta}(0, \xi) = 0.
\eqno{(\theequation{\mathit{e}, \mathit{f}})} \label{eq:bc_velocity}
$$
These boundary conditions are used to determine those for the streamfunction $\psi$ and vorticity  scalar $\omega$. It will be seen that the kinetic boundary condition (\ref{eq:bc_kinematic}), the zero normal velocity at the wall (\ref{eq:bc_velocity}{\it b}) and contact point (\ref{eq:bc_velocity}{\it c}), in addition to the zero normal velocity across the $z$-axis (\ref{eq:bc_velocity}{\it e}), shall be implicitly enforced by the boundary conditions for $\psi$. Moreover, the zero-stress boundary condition (\ref{eq:bc_dynamic}), together with the no-slip (zero-tangential velocity) at the wall (\ref{eq:bc_velocity}{\it a}) and symmetry condition at the $z$-axis (\ref{eq:bc_velocity}{\it f}) are enforced by the boundary conditions for $\omega$. The vorticity boundary condition at the contact point is derived following a special treatment.
 
\subsection{Boundary conditions for the streamfunction} 
\label{sec:bc_streamfunction}
From (\ref{eq:u_psi}{\it b}), boundary condition (\ref{eq:bc_velocity}{\it b}) implies that $\partial \psi/\partial \eta = 0$. Hence, $\psi$ is constant along the wall. The streamline value along the wall may be arbitrarily set to zero, i.e. $\psi(\eta, 0) = 0$. Following the same argument, (\ref{eq:bc_velocity}{\it c}) implies that the streamfunction value at the contact point is also zero: $\psi(\infty, \xi) = 0$.

From (\ref{eq:u_psi}{\it a}) and (\ref{eq:bc_kinematic}), the streamfunction at the bubble boundary must satisfy
\begin{equation} \label{eq:psi_sdiff}
    \frac{\partial \psi}{\partial \eta} = \hti^2\htiphi \xi' \quad \mbox{on} \quad \xi = 1.
\end{equation}
Making use of the definitions for $\hti$, $\htiphi$ and $\xi'$ provided in (\ref{eq:scale_factors})  and (\ref{eq:eta'_xi'}{\it b}), analytical integration of (\ref{eq:psi_sdiff}) results in 
\begin{equation} \label{eq:psi_s}
\psi(\eta,1) = -\frac{2 a^2 a'}{\eta^2+1} + f(\xi=1),
\end{equation}
where $f(\xi)$ accounts for the unknown function in $\xi$ that would be obtained had we been able to integrate $\partial \psi/\partial \xi$. Nevertheless, we can easily determine the value of the constant $ f(\xi=1)$ from the boundary condition $\psi(\infty,1) = 0$. This, of course, yields $f(\xi=1) = 0$.

Finally, (\ref{eq:bc_velocity}{\it e}) implies that  $\partial \psi/\partial \xi = 0$ and therefore $\psi(0,\xi)$ must be constant. Its value is found simply by evaluating (\ref{eq:psi_s}) on $\eta = 0$. 

\subsection{Boundary conditions for vorticity}
\label{sec:bc_vorticity}
From (\ref{eq:vort_def}), the velocity components are related to the vorticity $\omega$ through
\begin{equation}\label{eq:omega_u}
\omega = \frac{1}{\hti^2}\left[\frac{\partial}{\partial \eta}(\hti u_\xi) - 
\frac{\partial}{\partial \xi}(\hti u_\eta) \right].
\end{equation}
The boundary condition for the vorticity generated at a free surface may be conveniently expressed in terms of the tangential and normal velocity components $u_\eta(\eta, \xi, \tau)$ and $u_\xi(\eta, \xi, \tau)$ \citep{lundgren1999}.
Entering the zero-stress boundary condition (\ref{eq:bc_dynamic}) into (\ref{eq:omega_u}) leads to the following expression for the vorticity at the interface:
\begin{equation}
\omega = \frac{2}{\hti}\frac{\partial u_\xi}{\partial \eta} + \frac{2}{a} u_\eta \quad \mbox{on} \quad \xi = 1.
\end{equation}
An analytical expression of the first term may alternatively be obtained directly through (\ref{eq:bc_kinematic}). This results in (\ref{eq:bc_vorticity}{\it b}).
At the wall, the Dirichlet-type vorticity boundary condition for no-slip 
 \begin{equation}
\omega = -\frac{1}{\hti^2\htiphi}\frac{\partial^2 \psi}{\partial \xi^2} = 
-\frac{1}{\hti}\frac{\partial u_\eta}{\partial \xi}
\quad \mbox{on} \quad \xi = 0
\end{equation}
is usually used. The term $\partial^2 \psi / \partial \xi^2$ can be easily expressed in discretised form appealing to Thom's formula  \citep{thom1933} or any of its variants \citep{e1996}.
However, we must bear in mind that $\hti^2\htiphi \rightarrow 0$ as $\eta \rightarrow \infty$, i.e. our coordinate system is singular as it approaches the contact point. As a result, a Dirichlet-type boundary condition at the wall was found to be highly unstable.

\cite{takemura1996} employed the zero vorticity gradient condition $\partial \omega / \partial \xi = 0$. The zero vorticity gradient assumes a zero pressure gradient along the wall and consequently there is no vorticity generation at the wall \citep{lighthill1963}. In other words, the no-slip condition (\ref{eq:bc_velocity}{\it a}) is not enforced. 

The no-slip boundary condition may be alternatively imposed by Lighthill's dynamic description of vorticity. The idea is that the spurious non-zero slip velocity at the wall, i.e. $u_\eta (\eta, 0) \neq 0$, obtained by numerically differentiating $\psi$, should be counteracted by artificially creating vorticity on the wall. \cite{koumoutsakos1994} derived a Neumann-type vorticity boundary condition for no slip,
\begin{equation}
\nu \left.\frac{\partial \Omega}{\partial n}\right|_\mathrm{wall} = -\frac{U_t}{\Delta t},
\end{equation}
where $\partial / \partial n = \nhat \bcdot \bnabla$ denotes the directional derivative in the direction normal to the wall, $U_t$ is the spurious velocity tangential to the wall, and $\Delta t$ denotes a small interval of time (computational timestep) in the viscous timescale $t_v$ (cf. \S \ref{sec:timescales}) in which the vorticity flux is assumed constant.
In our dimensionless variables, this vorticity boundary condition becomes\begin{equation}
\frac{\partial \omega}{\partial \xi} = \frac{\hti \: u_\eta}{\Sc \: \Delta\tau_v} \quad \mbox{on} \quad \xi = 0,
\end{equation}
where $\Delta\tau_v$ is the viscous computational timestep (see \S \ref{sec:num_procedure}).

The geometry of the bubble very close to the contact line can be approximated by a two dimensional wedge with contact angle $\theta$ and polar coordinates ($r$, $\varphi$). In viscous corner flow, inertial effects may be neglected and Stokes momentum equation is described by the biharmonic equation $\nabla^4 \psi = 0$.

An approximate yet acceptable boundary condition for vorticity at the contact point ($\eta = \infty$) may be determined from the flow solution to the contact line pinning (CR mode) scenario.
The bubble surface is then taken as a hinged plane on $\varphi = \theta$ which rotates around the origin with angular velocity $\theta'$ on which the zero shear stress condition applies. The horizontal plane ($\varphi= 0$) is a solid wall at rest which is impermeable and allows no slip. Dimensional analysis gives a solution of the form \citep{gelderblom2012, moffatt1964}
\begin{equation}
\psi = \theta' r^2 f(\varphi),
\end{equation}
where $f(\varphi)$ is a suitable function. Hence, the vorticity close to the contact line must be independent of $r$, i.e. $\omega = \omega(\varphi)$.
As $\eta \rightarrow \infty$, it may be shown that $\varphi \rightarrow \theta\xi$, which leads to the zero vorticity gradient condition $\partial \omega / \partial \eta = 0$ across the contact line.

Finally, referring to (\ref{eq:bc_velocity}{\it e}) and (\ref{eq:bc_velocity}{\it f}) to evaluate (\ref{eq:omega_u}) on $\eta = 0$ results in a zero vorticity boundary condition (consistent with symmetry) on the $z$-axis. 

\section{Transformation matrix}
\label{sec:transformation_matrix}

Any vector expressed in Cartesian coordinates, $\bv_{\{x,y,z\}} = v_x \:\ex + v_y \:\ey + v_z \:\ez$, may be mapped to tangent-sphere coordinates $\bv_{\{\eta,\xi,\phi\}} = v_\eta \:\eeta + v_\xi \:\exi + v_\phi \:\ephi$ by 
\begin{equation}
\bv_{\{x,y,z\}}= \mathsfbi{J}\:\bv_{\{\eta,\xi,\phi\}},
\end{equation}
with
\begin{equation}
\setlength{\arraycolsep}{8pt}
\renewcommand{\arraystretch}{1.3}
\mathsfbi{J} = \left[
\begin{array}{ccc}
  \displaystyle -\frac{\eta^2-\xi^2}{\den}\cos\phi & \displaystyle -\frac{2\eta\xi}{\den}\cos\phi & -\sin\phi \\
  \displaystyle -\frac{\eta^2-\xi^2}{\den}\sin\phi & \displaystyle -\frac{2\eta\xi}{\den}\sin\phi & \cos\phi \\
  \displaystyle -\frac{2\eta\xi}{\den}             & \displaystyle \frac{\eta^2-\xi^2}{\den}      & 0 \\
\end{array}  \right] .
\end{equation}
Since $\mathsfbi{J}$ is orthogonal, then  $\bv_{\{\eta,\xi,\phi\}}= \mathsfbi{J^T}\:\bv_{\{x,y,z\}}$ gives the opposite transformation. This is useful for plotting purposes, since the Cartesian velocity components in the $x$-$z$ plane ($\phi = 0$) may be easily obtained from $u_\eta$ and $u_\xi$ as follows:
\refstepcounter{equation}
$$ 
u_x = -\frac{(\eta^2-\xi^2)u_\eta + 2\eta\xi \:u_\xi}{\den}, \qquad
u_z = \frac{(\eta^2-\xi^2)\:u_\xi - 2\eta\xi \:u_\eta}{\den}.
\eqno{(\theequation{\mathit{a}, \mathit{b}})}
$$

\bibliography{he2-references}

\begin{thebibliography}{29}
\expandafter\ifx\csname natexlab\endcsname\relax\def\natexlab#1{#1}\fi
\def\au#1{#1} \def\ed#1{#1} \def\yr#1{#1}\def\at#1{#1}\def\jt#1{\textit{#1}}
  \def\bt#1{#1}\def\bvol#1{\textbf{#1}} \def\vol#1{#1} \def\pg#1{#1}
  \def\publ#1{#1}\def\arxiv#1{#1}\def\org#1{#1}\def\st#1{\textit{#1}}

\bibitem[Bataller {\em et~al.\/}(2009)Bataller, Miqueu, Plantier, Daridon,
  Jaber, Abbasi, Saghir \& Bou-Ali]{bataller2009}
{\sc \au{Bataller, H.}, \au{Miqueu, C.}, \au{Plantier, F.}, \au{Daridon,
  J.-L.}, \au{Jaber, T.~J.}, \au{Abbasi, A.}, \au{Saghir, M.~Z.} \&
  \au{Bou-Ali, M.~M.}} \yr{2009}  \at{Comparison between experimental and
  theoretical estimations of the thermal expansion, concentration expansion
  coefficients, and viscosity for binary mixtures under pressures up to 20
  {MP}a}.  \jt{J. Chem. Eng. Data}  \bvol{54}~(6),  \pg{1710--1715}.

\bibitem[Batchelor(1979)]{batchelor1979}
{\sc \au{Batchelor, G.~K.}} \yr{1979}  \at{Mass transfer from a particle
  suspended in fluid with a steady linear ambient velocity distribution}.
  \jt{J. Fluid Mech.}  \bvol{95},  \pg{369--400}.

\bibitem[Chu \& Prosperetti(2016{\natexlab{{\em a\/}}})]{chu2016}
{\sc \au{Chu, S.} \& \au{Prosperetti, A.}} \yr{2016{\natexlab{{\em a\/}}}}
  \at{Dissolution and growth of a multicomponent drop in an immiscible liquid}.
   \jt{J. Fluid Mech.}  \bvol{798},  \pg{787--811}.

\bibitem[Chu \& Prosperetti(2016{\natexlab{{\em b\/}}})]{chu2016.1}
{\sc \au{Chu, S.} \& \au{Prosperetti, A.}} \yr{2016{\natexlab{{\em b\/}}}}
  \at{History effects on the gas exchange between a bubble and a liquid}.
  \jt{Phys. Rev. Fluids}  \bvol{1}~(064202),  \pg{1--20}.

\bibitem[Dietrich {\em et~al.\/}(2016)Dietrich, Wildeman, Visser, Hofhuis,
  Kooij, Zandvliet \& Lohse]{dietrich2016}
{\sc \au{Dietrich, E.}, \au{Wildeman, S.}, \au{Visser, C.~W.}, \au{Hofhuis,
  K.}, \au{Kooij, E.~S.}, \au{Zandvliet, H. J.~W.} \& \au{Lohse, D.}} \yr{2016}
   \at{Role of natural convection in the dissolution of sessile droplets}.
  \jt{J. Fluid Mech.}  \bvol{794},  \pg{45--67}.

\bibitem[Enr\'iquez {\em et~al.\/}(2013)Enr\'iquez, Hummelink, Bruggert, Lohse,
  Prosperetti, van~der Meer \& Sun]{enriquez2013}
{\sc \au{Enr\'iquez, O.~R.}, \au{Hummelink, C.}, \au{Bruggert, G.-W.},
  \au{Lohse, D.}, \au{Prosperetti, A.}, \au{van~der Meer, D.} \& \au{Sun, C.}}
  \yr{2013}  \at{Growing bubbles in a slightly supersaturated liquid solution}.
   \jt{Rev. Sci. Instrum.}  \bvol{84}~(065111).

\bibitem[Enr\'iquez {\em et~al.\/}(2014)Enr\'iquez, Sun, Lohse, Prosperetti \&
  van~der Meer]{enriquez2014}
{\sc \au{Enr\'iquez, O.~R.}, \au{Sun, C.}, \au{Lohse, D.}, \au{Prosperetti, A.}
  \& \au{van~der Meer, D.}} \yr{2014}  \at{The quasi-static growth of
  $\mathrm{CO}_2$ bubbles}.  \jt{J. Fluid Mech.}  \bvol{741}~(R1).

\bibitem[Epstein \& Plesset(1950)]{epstein1950}
{\sc \au{Epstein, P.~S.} \& \au{Plesset, M.~S.}} \yr{1950}  \at{On the
  stability of gas bubbles in liquid-gas solutions}.  \jt{J. Chem. Phys.}
  \bvol{18}~(11),  \pg{1505--1509}.

\bibitem[Gelderblom {\em et~al.\/}(2012)Gelderblom, Bloemen \&
  Snoeijer]{gelderblom2012}
{\sc \au{Gelderblom, H.}, \au{Bloemen, O.} \& \au{Snoeijer, J.~H.}} \yr{2012}
  \at{Stokes flow near the contact line of an evaporating drop}.  \jt{J. Fluid
  Mech.}  \bvol{709},  \pg{69--84}.

\bibitem[Harvey {\em et~al.\/}(2005)Harvey, Kaplan \& Burnett]{harvey2005}
{\sc \au{Harvey, A.~H.}, \au{Kaplan, S.~G.} \& \au{Burnett, J.~H.}} \yr{2005}
  \at{Effect of dissolved air on the density and refractive index of water}.
  \jt{Int. J. Thermophys.}  \bvol{26}~(5),  \pg{1495--1514}.

\bibitem[Koumoutsakos {\em et~al.\/}(1994)Koumoutsakos, Leonard \&
  P\'epin]{koumoutsakos1994}
{\sc \au{Koumoutsakos, P.}, \au{Leonard, A.} \& \au{P\'epin, F.}} \yr{1994}
  \at{Boundary conditions for viscous vortex methods}.  \jt{J. Comput. Phys.}
  \bvol{113}~(1),  \pg{52--61}.

\bibitem[Lighthill(1963)]{lighthill1963}
{\sc \au{Lighthill, M.~J.}} \at{ \yr{1963} } \bt{In {\em Laminar boundary
  layers\/} (ed. \ed{L.~Rosenhead})},  \pg{pp. 46--57}.  \publ{Oxford
  University Press}.

\bibitem[Lohse(2016)]{lohse2016}
{\sc \au{Lohse, D.}} \yr{2016}  \at{Towards controlled liquid-liquid
  microextraction}.  \jt{J. Fluid Mech.}  \bvol{84},  \pg{1--4}.

\bibitem[Lundgren \& Koumoutsakos(1999)]{lundgren1999}
{\sc \au{Lundgren, T.} \& \au{Koumoutsakos, P.}} \yr{1999}  \at{On the
  generation of vorticity at a free surface}.  \jt{J. Fluid Mech.}  \bvol{382},
   \pg{351--366}.

\bibitem[Michaelides(2003)]{michaelides2003}
{\sc \au{Michaelides, E.~E.}} \yr{2003}  \at{Hydrodynamic force and heat/mass
  transfer from particles, bubbles, and drops - {T}he {F}reeman scholar
  lecture}.  \jt{J. Fluid Eng.}  \bvol{125},  \pg{209--238}.

\bibitem[Moffatt(1964)]{moffatt1964}
{\sc \au{Moffatt, H.~K.}} \yr{1964}  \at{Viscous and resistive eddies near a
  sharp corner}.  \jt{J. Fluid Mech.}  \bvol{18},  \pg{1--18}.

\bibitem[Moon \& Spencer(1988)]{moon1988}
{\sc \au{Moon, P.} \& \au{Spencer, D.~E.}} \yr{1988} {\em Field theory handbook
  - including coordinate systems, differential equations and their
  solutions\/}, 2nd edn.,  \pg{p. 104}.  \publ{Springer-Verlag}.

\bibitem[Pe{\~n}as-L\'opez {\em et~al.\/}(2016{\natexlab{{\em
  a\/}}})Pe{\~n}as-L\'opez, van Elburg, Parrales \&
  Rodr\'iguez-Rodr\'iguez]{PenasLopez_etalPRF2016}
{\sc \au{Pe{\~n}as-L\'opez, P.}, \au{van Elburg, B.}, \au{Parrales, M.~A.} \&
  \au{Rodr\'iguez-Rodr\'iguez, J.}} \yr{2016{\natexlab{{\em a\/}}}}
  \at{Diffusion of dissolved $\mathrm{CO}_2$ in water propagating from a
  cylindrical bubble in a horizontal {H}ele-{S}haw cell}.  \jt{Submitted to
  Phys. Rev. Fluids} .

\bibitem[Pe{\~n}as-L\'opez {\em et~al.\/}(2015)Pe{\~n}as-L\'opez, Parrales \&
  Rodr\'iguez-Rodr\'iguez]{penas2015}
{\sc \au{Pe{\~n}as-L\'opez, P.}, \au{Parrales, M.~A.} \&
  \au{Rodr\'iguez-Rodr\'iguez, J.}} \yr{2015}  \at{Dissolution of a
  $\mathrm{CO_2}$ spherical cap bubble adhered to a flat surface in
  air-saturated water}.  \jt{J. Fluid Mech.}  \bvol{775},  \pg{53--76}.

\bibitem[Pe{\~n}as-L\'opez {\em et~al.\/}(2016{\natexlab{{\em
  b\/}}})Pe{\~n}as-L\'opez, Parrales, Rodr\'iguez-Rodr\'iguez \& van~der
  Meer]{penas2016}
{\sc \au{Pe{\~n}as-L\'opez, P.}, \au{Parrales, M.~A.},
  \au{Rodr\'iguez-Rodr\'iguez, J.} \& \au{van~der Meer, D.}}
  \yr{2016{\natexlab{{\em b\/}}}}  \at{The history effect in bubble growth and
  dissolution. {P}art 1. {T}heory}.  \jt{J. Fluid Mech.}  \bvol{800},
  \pg{180--212}.

\bibitem[Rosner \& Epstein(1972)]{rosner1972}
{\sc \au{Rosner, D.~E.} \& \au{Epstein, M.}} \yr{1972}  \at{Effects of
  interface kinetics, capillarity and solute diffusion on bubble growth rates
  in highly supersaturated liquids}.  \jt{Chem. Eng. Sci.}  \bvol{27}~(1),
  \pg{69--88}.

\bibitem[Scriven(1959)]{scriven1959}
{\sc \au{Scriven, L.~E.}} \yr{1959}  \at{On the dynamics of phase growth}.
  \jt{Chem. Eng. Sci.}  \bvol{10}~(1),  \pg{1--13}.

\bibitem[Shim {\em et~al.\/}(2014)Shim, Wan, Hilgenfeldt, Panchal \&
  Stone]{shim2014}
{\sc \au{Shim, S.}, \au{Wan, J.}, \au{Hilgenfeldt, S.}, \au{Panchal, P.~D.} \&
  \au{Stone, H.~A.}} \yr{2014}  \at{Dissolution without disappearing:
  multicomponent gas exchange for $\mathrm{CO_2}$ bubbles in a microfluidic
  channel}.  \jt{Lab Chip}  \bvol{14},  \pg{2428--2436}.

\bibitem[Sun \& Cubaud(2011)]{sun2011}
{\sc \au{Sun, R.} \& \au{Cubaud, T.}} \yr{2011}  \at{Dissolution of carbon
  dioxide bubbles and microfluidic multiphase flows}.  \jt{Lab Chip}
  \bvol{11},  \pg{2924--2928}.

\bibitem[Szekely \& Martins(1971)]{szekely1971}
{\sc \au{Szekely, J.} \& \au{Martins, G.~P.}} \yr{1971}  \at{Non equilibrium
  effects in the growth of spherical gas bubbles due to solute diffusion}.
  \jt{Chem. Eng. Sci.}  \bvol{26}~(1),  \pg{147--159}.

\bibitem[Takemura {\em et~al.\/}(1996)Takemura, Liu \& Yabe]{takemura1996}
{\sc \au{Takemura, F.}, \au{Liu, Q.} \& \au{Yabe, A.}} \yr{1996}  \at{Effect of
  density-induced natural convection on the absorption process of single
  bubbles under a plate}.  \jt{Chem. Eng. Sci.}  \bvol{51}~(20),
  \pg{4551--4560}.

\bibitem[Thom(1933)]{thom1933}
{\sc \au{Thom, A.}} \yr{1933}  \at{The flow past circular cylinders at low
  speeds}.  \jt{Proc. R. Soc. A}  \bvol{141}~(845),  \pg{651--669}.

\bibitem[Volk {\em et~al.\/}(2015)Volk, Rossi, K{\"a}hler, Hilgenfeldt \&
  Mar\'in]{volk2015}
{\sc \au{Volk, A.}, \au{Rossi, M.}, \au{K{\"a}hler, C.~J.}, \au{Hilgenfeldt,
  S.} \& \au{Mar\'in, A.}} \yr{2015}  \at{Growth control of sessile
  microbubbles in {PDMS} devices}.  \jt{Lab Chip}  \bvol{15},  \pg{4607--4613}.

\bibitem[Weinan \& Liu(1996)]{e1996}
{\sc \au{Weinan, E.} \& \au{Liu, J.-G.}} \yr{1996}  \at{Vorticity boundary
  condition and related issues for finite difference schemes}.  \jt{J. Comput.
  Phys.}  \bvol{124}~(2),  \pg{368--382}.

\end{thebibliography}
\bibliographystyle{jfm}

\end{document}